\documentclass[a4paper]{article}
\usepackage{stmaryrd} % gives \llparenthesis, \rrparenthesis, \llceil, \rrceil
\usepackage[lutzsyntax]{virginialake}\vlsmallbrackets\aftrianglefalse
\usepackage{proof}
\usepackage{amssymb}
\usepackage{amsmath}
\usepackage{amsthm} %% not compatible with llncs
\usepackage{latexsym}
\usepackage{paralist}
\usepackage{calrsfs}
\usepackage{hyperref}
\usepackage{graphicx}
\usepackage{txfonts} % supplies \multimapdots
\usepackage{color} %usage \textcolor{nameofhecolor}{texttocolor}
\usepackage{subfigure} 
\usepackage{textcomp}
\newcommand{\Set}[1]{ \{ #1 \}}
\newcommand{\Size}[1]{|#1|}

\newcommand{\smalltitle}[1]{\small #1} 
\newcommand{\ST}{such that}

\newcommand{\llcxlinlamcalc}{linear lambda calculus}
\newcommand{\llcxLinlamcalc}{Linear lambda calculus}
\newcommand{\llcxwithexs}{with explicit substitutions}
\newcommand{\llcxlamcalc}{lambda calculus}
\newcommand{\llcxLamcalc}{Lambda calculus}
\newcommand{\llcxlinlamterm}{linear lambda term}
\newcommand{\llcxlamterm}{lambda term}

\newcommand{\llcxlinlamterms}{\llcxlinlamterm s}
\newcommand{\llcxlamterms}{\llcxlamterm s}

\newcommand{\llcxlamvar}{lambda variable}
\newcommand{\llcxlamvars}{\llcxlamvar s}
\newcommand{\IFF}{iff}
\newcommand{\ie}{i.e.}
\newcommand{\wrt}{w.r.t.}
\newcommand{\dfn}[1]{\emph{#1}}
\newcommand{\grammareq}{::=\ }

\newcommand{\PTIMEC}{\mathsf{ptime}\mbox{-complete}}
\newcommand{\NPTIMEC}{\mathsf{nptime}\mbox{-complete}}
\newcommand{\DI}{DI}
\newcommand{\NEL}{\mathsf{NEL}}
\newcommand{\BV}{\mathsf{BV}}
\newcommand{\SBV}{\mathsf{SBV}}

\newcommand{\SBVsub}{\mathsf{B}}
\newcommand{\BVT}{\BV\mathtt{r}}
\newcommand{\SBVT}{\SBV\mathtt{r}}

\newcommand{\CCS}{\mathsf{CCS}}
\newcommand{\MLL}{\mathsf{MLL}}
\newcommand{\IMLL}{\mathsf{IMLL}}

\definecolor{blue}{rgb}{0.2,0.2,0.4}
\definecolor{brown}{rgb}{0.7,0.4,0.3}
\definecolor{red}{rgb}{1,0.2,0}
\definecolor{yellow}{rgb}{1,1,0}

\theoremstyle{plain}
\newtheorem{theorem}{Theorem}[section]
\newtheorem{lemma}[theorem]{Lemma}
\newtheorem{proposition}[theorem]{Proposition}
\newtheorem{corollary}[theorem]{Corollary}

\theoremstyle{definition}

\newtheorem{remark}[theorem]{Remark}

\newcommand{\llcxVar}{\mathcal{V}} %
\newcommand{\llcxSet}[1]{\Lambda_{#1}} %
\newcommand{\llcxFV}[1]{\operatorname{fv}(#1)} % Free variables set
\newcommand{\llcxF}[2]{\lambda #1.#2} % lambda abstraction
\newcommand{\llcxA}[2]{(#1)\, #2} % application
\newcommand{\llcxE}[3]{(#1)\, \Set{#2 = #3}} % Explicit subst. #3 for #2 in #1
\newcommand{\llcxM}{M} % lambda terms names
\newcommand{\llcxN}{N}
\newcommand{\llcxP}{P}
\newcommand{\llcxQ}{Q}
\newcommand{\llcxX}{x} % lambda variables names
\newcommand{\llcxY}{y}
\newcommand{\llcxW}{w}
\newcommand{\llcxZ}{z}
\newcommand{\llcxo}{o} % lambda variables channel names
\newcommand{\llcxno}{\overline{o}} % lambda variables channel names
\newcommand{\llcxp}{p} %
\newcommand{\llcxpP}{p'} %
\newcommand{\llcxq}{q} %
\newcommand{\llcxnp}{\overline{p}} %

\newcommand{\llcxBeta}{\rightarrow} % Rewriting relations 
\newcommand{\llcxSOSJud}[2]{ #1 \Rightarrow #2} %

\newcommand{\llcxSOSBrule  }{\mathrm{lft}} % Rewriting rule names
\newcommand{\llcxSOStrarule}{\mathrm{tra}} %
\newcommand{\llcxSOSrflrule}{\mathrm{rfl}} %
\newcommand{\llcxSOSfrule  }{\mathrm{f}} %
\newcommand{\llcxSOSalrule}{\mathrm{@l}} %
\newcommand{\llcxSOSarrule}{\mathrm{@r}} %
\newcommand{\llcxSOSslrule}{\sigma\mathrm{l}} %
\newcommand{\llcxSOSsrrule}{\sigma\mathrm{r}} %

\newcommand{\vlrulename}[1]{\hbox{$\vlleftlabel {#1}$\!}} %
\vllineartrue
\renewcommand{\vlscn}[1]{
\!\ifvirginialakesmallbrackets\{\else\left\{\fi #1
  \ifvirginialakesmallbrackets\}\else\right\}\fi   } % context
\newcommand{\vlone}{\circ} % multiplicative unit
\newcommand{\vlfo}[2]{\vlqu{#1}{#2}}
\newcommand{\vlex}[2]{\vlqu{#1}{#2}}

\newcommand{\vlpl}{\vlor}
\renewcommand{\vlne}[1]{\overline{#1}}
\newcommand{\strFN}[1]{\operatorname{FN}(#1)} % free names in structures
\newcommand{\strK}{K} % structure names
\newcommand{\strP}{P} 
\newcommand{\strR}{R} 
\newcommand{\strS}{S}
\newcommand{\strT}{T}
\newcommand{\strU}{U}
\newcommand{\strV}{V}
\newcommand{\nstrR}{\vlne\strR} 

\newcommand{\nstrT}{\vlne\strT}

\newcommand{\atma}{a} % atoms
\newcommand{\atmb}{b}
\newcommand{\atmc}{c}

\newcommand{\natma}{\vlne\atma} % negated atoms
\newcommand{\natmb}{\vlne\atmb}
\newcommand{\natmc}{\vlne\atmc}

\newcommand{\subst}[2]{\{^{#1}\!/\!_{#2}\}} % substitutes #1 for #2
\newcommand{\bvtrdrule}{\mathrm{r}\downarrow}
\newcommand{\bvtrurule}{\mathrm{r}\uparrow}

\newcommand{\bvtrhorule}{\rho}
\newcommand{\bvtmixprule}{\mathrm{mixp}} % derivable rule names
\newcommand{\bvtpmixrule}{\mathrm{pmix}} 
\newcommand{\bvtdefdrule}{\mathrm{def}\downarrow}
\newcommand{\bvtdefurule}{\mathrm{def}\uparrow}
\newcommand{\bvtintdrule}{\mathrm{i}\downarrow}
\newcommand{\bvtinturule}{\mathrm{i}\uparrow} 
\newcommand{\bvtseqdrule}{\mathrm{q}\downarrow} 
\newcommand{\bvtsequrule}{\mathrm{q}\uparrow} 
\newcommand{\bvtatidrule}{\mathrm{ai}\downarrow} 
\newcommand{\bvtatiurule}{\mathrm{ai}\uparrow} 
\newcommand{\bvtswirule }{\mathrm{s}} 
\newcommand{\bvtpludrule}{\mathrm{p}\downarrow}
\newcommand{\bvtpluurule}{\mathrm{p}\uparrow}

\newcommand{\bvtorerule}{\mathrm{o-ren}}
\newcommand{\bvtsinrule}{\mathrm{s-intro}}
\newcommand{\bvtslrule }{\mathrm{s-}\lambda}
\newcommand{\bvtsalrule}{\mathrm{s-}@l}
\newcommand{\bvtsarrule}{\mathrm{s-}@r}
\newcommand{\bvtsvarule}{\mathrm{s-}var}
\newcommand{\bvtDder}{\mathcal{D}} % derivation names
\newcommand{\bvtPder}{\mathcal{P}}
\newcommand{\bvtInfer}[2]{#1: #2}
\newcommand{\bvtJudGen}[2]{\vdash_{#1 % system of reference
}^{#2 % subset of rules
}}
\newcommand{\bvtJud}[1]{\vdash_{\BVT}^{#1 % subset of rules
}}
\newcommand{\bvtJ}{\vdash}

\newcommand{\mllPder}{\Pi} % derivation names
\newcommand{\mllPderA}{\Pi_A} % derivation names
\newcommand{\mllPderB}{\Pi_B} % derivation names
\newcommand{\mllA}{A} % formulas names
\newcommand{\mllB}{B} %
\newcommand{\mllnA}{\vlne{\mllA}} % formulas names
\newcommand{\mllVarA}{\alpha} % variable names
\newcommand{\mllVarB}{\beta}
\newcommand{\mllImpl}{\multimap}
\newcommand{\mllTens}{\otimes}

\newcommand{\mllax  }{\mathrm{ax}} % rule names
\newcommand{\mllcut }{\mathrm{cut}}
\newcommand{\mlltens}{\mllTens}
\newcommand{\mllimpl}{\mllImpl}
\newcommand{\mllHasType}{\!:\!}

\newcommand{\mllJud}{\vdash_{\MLL}}
\newcommand{\mapMLLToBVT}[1]{ #1^\bullet} %
\newcommand{\imllJud}{\vdash_{\IMLL}}

\newcommand{\mapLcToDi}[2]
           {\llparenthesis\,\! #1\, \rrparenthesis_{#2}} % #2 is out channel
\newcommand{\atmo}{o}
\newcommand{\atmp}{p}
\newcommand{\atmq}{q}
\newcommand{\natmo}{\vlne\atmo}

\begin{document}

\newcommand{\TITLE}{\llcxLinlamcalc\ \llcxwithexs\ 
\\ as proof-search in Deep Inference}
\newcommand{\RUNTITLE}{\llcxLinlamcalc\ \llcxwithexs\
as proof-search in Deep Inference}
\title{\TITLE}

\author{
  \large Luca Roversi%
  \\[1ex]
    Dip. di Informatica -- Univ. di Torino\\
    {\url{http://www.di.unito.it/~rover/}}\\[3ex]
%   \and
%   \large Another author%
%   \\[1ex]
%     University of  ..., address \\
%     {\url{}}
}

\date{\today}
\maketitle

\pagestyle{myheadings}
\markboth{Roversi}{\TITLE}

\begin{abstract}
$\SBV$  is a deep inference system that extends the set of logical operators of
multiplicative linear logic with the non commutative operator \emph{seq}.
\par
We introduce the logical system $\SBVT$ which extends $\SBV$ by adding a
self-dual atom-renaming operator to it. We prove that the cut elimination holds
on $\SBVT$.
\par
$\SBVT$ and its cut free subsystem $\BVT$ are complete and sound with respect to
linear \llcxLamcalc\ \llcxwithexs. Under any strategy, a sequence of evaluation
steps of any linear $\lambda$-term $\llcxM$ becomes a process of proof-search in
$\SBVT$ ($\BVT$) once $\llcxM$ is mapped into a formula of $\SBVT$.
\par
Completeness and soundness follow from simulating linear $\beta$-reduction
\llcxwithexs\ as processes. The role of the new renaming operator of $\SBVT$ is
to rename channel-names on-demand. This simulates the substitution that occurs
in a $\beta$-reduction.
\par
Despite $\SBVT$ is a minimal extension of $\SBV$ its proof-search can compute
all boolean functions, as linear \llcxlamcalc\ \llcxwithexs\ can compute all
boolean functions as well. So, proof search of $\SBVT$ and $\BVT$ is at least
$\PTIMEC$.
\end{abstract}

%\tableofcontents

%%%%%%%%%%%%%%%%%%%%%%%%%%%%%%%%%%%%%%%%%%%%%%%%%%%%%%
\section{Introduction}
\label{section:Introduction}
We shall see how the functional computation that \llcxlamcalc\ \llcxwithexs\
develops relates to proof-search inside an extension of $\SBV$
\cite{Gugl:06:A-System:kl}, the system at the core of \emph{deep inference}
(\DI).
%%%%%%%%%%%%%%%%%%%%%
\paragraph{System $\SBV$.} Semantic motivation, intuitions, examples of its use
and a cut elimination theorem of the system $\SBV$ are in
\cite{Gugl:06:A-System:kl}. The cut free sub-system of $\SBV$ is $\BV$. The idea
leading to $\SBV$ is that the logical systems we may rely formal reasoning on
must not necessarily exploit \emph{shallow} rules, as opposed to \emph{deep}
ones. Rules of sequent and natural deduction systems are shallow because they
build proofs with a form that mimic the structure of the formula they prove.
Deep rules, instead, apply arbitrarily deep in the tree representation of a
formula. Thanks to the above deepness, $\BV$ substantially extends
multiplicative linear logic ($\MLL$) \cite{GirardTaylorLafont89} with
``$\vlse$'', the non commutative binary operator \emph{seq}. Many sources of
evidence about the relevance of $\BV$ exist. The deep application of rules in
$\BV$ is strictly connected to its expressiveness, as compared to $\MLL$. Any
limits we might put on the application depth of $\BV$ rules would yield a
strictly less expressive system \cite{Tiu:06:A-System:ai}. Moreover, under the
analogy ``processes-as-formulas and communication-as-proof-search'',
\cite{Brus:02:A-Purely:wd} shows that the operator seq models the sequential
behavior $\CCS$, the system of concurrent and communicating processes
\cite{Miln:89:Communic:qo}. Also, $\BV$, which is $\NPTIMEC$
\cite{Kahr:07:System-B:fk}, has then been extended with linear logic 
exponentials, in the system $\NEL$
\cite{GuglStra:01:Non-comm:rp,GuglStra:02:A-Non-co:lq,GuglStra:02:A-Non-co:dq,
StraGugl:09:A-System:vn} , whose provability is undecidable
\cite{Stra:03:System-N:mb}. Finally, strong connections between 
$\BV$ develops and the evolution of discrete quantum systems are emerging
\cite{BlutPanaStra:08:The-Logi:lh,BlutGuglIvanPana:10:A-Logica:uq}
%%%%%%%%%%%%%%%%%%
\paragraph{\llcxLinlamcalc\ \llcxwithexs.}
There is a vast literature on \emph{explicit substitutions}. We just recall
\cite{AbadiCCL91,Lescanne94:POPL94,Mellies:TLCA95,Rose:BRICS-LS-96-3} as
pointers. We focus on the simplest version of \llcxlamcalc\ endowed with the
obvious notion of explicit substitutions which embodies the kernel of functional
programming at its simplest level. The functions \llcxlinlamcalc\ \llcxwithexs\
represents use their arguments exactly once in the course of the evaluation. The
set of functions we can express in it are quite limited, but ``large'' enough to
let the decision about which is the normal form of its \llcxlamterms\ a
\emph{polynomial time complete} problem \cite{Mairson:2003JFP}, if we take the
polynomial time Turing machines as computational model of reference, of course.
Recall that ``\emph{\llcxwithexs}'' means that operation substituting a
\llcxlamterm\ for a \llcxlamvar, in the course of a $\beta$-reduction is not
meta, but a syntactical construction.
%%%%%%%%%%%%%%%%%%
\paragraph{Leading motivations.}
Our motivation is to search how structural proof theory, based on \DI\
methodology, can contribute to paradigmatic programming language design. The
reason why we think \DI\ can be useful to this respect is that structural proof
theory of a quite vast range of logics has become very regular and modular.
Proof theory of \DI\ is now developed for classical
\cite{Brun:03:Atomic-C:oz,Brun:06:Cut-Elim:cq,Brun:06:Deep-Inf:qy,
Brun:06:Locality:zh, BrunTiu:01:A-Local-:mz}, intuitionistic
\cite{Tiu:06:A-Local-:gf}, linear
\cite{Stra:02:A-Local-:ul,Stra:03:Linear-L:lp,Stra:03:MELL-in-:oy,
Di-G:04:Structur:wy} and modal
\cite{Brun::Deep-Seq:ay,GoreTiu:06:Classica:uq,Stou:06:A-Deep-I:rt} logics,
indeed.
\par
We expect that much regularity and modularity at the proof-theory level can
highlight useful inherent properties and new primitives, or evaluation
strategies, at the level of programs. The point is to look for the
computational interpretation of derivations in \DI\ style, in the same vein as
the one we are used to with shallow inference. For example, a source of new
programming primitives, or evaluation strategies, can be \DI\ deductive systems
whose inference rules only manipulate atoms of formulas, and for which new
notions of proof normalization exist, in addition to cut elimination.
%%%%%%%%%%%%%%%%%%
\paragraph{Starting observation.}
A typical way to illustrate the properties of $\BV$ is to show that any
derivation of the sequent $\mllJud \mllA_1,\ldots,\mllA_m$ of $\MLL$ embeds into
a derivation of $\BV$ under $\mapMLLToBVT{(\cdot)}$ that maps par and tensor of
$\MLL$ into par and copar of $\BV$, respectively, and whose extension to $\MLL$
sequents is:
\begin{align}
\label{align:mllax-into-par-tens}
\mapMLLToBVT{(\mllJud \mllA_1,\ldots,\mllA_m)} & =
\vlsbr[\mapMLLToBVT{\mllA_1}\vlpa\vldots\vlpa\mapMLLToBVT{\mllA_m}]
\end{align}
However, alternatively to \eqref{align:mllax-into-par-tens},
\emph{intuitionistic} multiplicative linear logic ($\IMLL$) can embed into $\BV$
by mapping sequents of $\IMLL$ into formulas of $\SBV$:
\vlstore{%
\vlsbr<(\vlne{\mapMLLToBVT{\mllVarA_1}}
       ;\vldots;
        \vlne{\mapMLLToBVT{\mllVarA_m}})
      ;\mapMLLToBVT{\mllVarB}>
}
\begin{align}
\label{align:mllax-into-seq}
\mapMLLToBVT{(\mllVarA_1,\ldots,\mllVarA_m\imllJud\mllVarB)} & =
\vlread
\end{align}
After~\eqref{align:mllax-into-seq}, a first step is recalling that every axiom
$\mllA\imllJud\mllA$ can give a type to a variable $\llcxX$ of \llcxlinlamcalc\
as in $\llcxX\mllHasType\mllA\imllJud\llcxX\mllHasType\mllA$. A second step is
recalling the intuition behind the interpretation of any structure
$\vlsbr<\strR;\strT>$ of $\BV$. The atoms of $\strR$, and $\strT$ will never
interact. So, the following representation $\mapLcToDi{\llcxX}{\atmo}$ of
$\llcxX$ as structure in $\BV$ can make sense:
\vlstore{\vlsbr<\llcxX;\natmo>}
\begin{align}
\label{align:llcx-into-bv}
\mapLcToDi{\llcxX}{\atmo} & = \vlread
\end{align}
In~\eqref{align:llcx-into-bv} $\llcxX$ becomes the name of the input channel to
the left of $\vlse$ that will eventually be forwarded to the output channel
$\atmo$, associated to $\llcxX$ by $\vlse$.
Noticeably, \eqref{align:llcx-into-bv} strongly resembles the base clause:
\begin{align}
\label{align:lc-into-pi-base}
\vlstore{{\vls\llcxX(\!\vlone\!)\,\vldot\,\natmo\langle\!\vlone\!\rangle}}
\textlbrackdbl\llcxX\textrbrackdbl_\atmo & = \vlread
\end{align}
\noindent
of the, so called, \emph{output-based} embedding of the \emph{standard}
\llcxlamcalc\ \emph{\llcxwithexs} into $\pi$-calculus
\cite{DBLP:conf/concur/BakelV09}. In it, ``$\vldot$'' is the sequential
composition of the $\pi$-calculus and $\vlone$ a generic, essentially
place-holder, variable. The whole structure is a \emph{forwarder}, in
accordance with the terminology of \cite{HondaYoshida:TCS1995}. We recall from
\cite{DBLP:conf/concur/BakelV09} that \emph{output-based} embedding is more
liberal than the more popular \emph{input-based} embeddings, inspired to the one
in \cite{DBLP:journals/mscs/Milner92}. Output-based one simulates \emph{spine}
reduction of \emph{standard} \llcxlamcalc\ \llcxwithexs, while the input-based
embedding simulates \emph{lazy} $\beta$-reduction strategy.
%%%%%%%%%%%%%%%%%
\paragraph{The need to extend $\BV$.}
The essential correspondence between \eqref{align:llcx-into-bv}, and
\eqref{align:lc-into-pi-base} rise the question about how could we represent,
\emph{at least a fragment} of standard \llcxlamcalc\ as a process of
proof-search inside $\BV$, in the style of the above output-based embedding. The
main missing ingredient is what we can dub as \emph{on-the-fly renaming} of
channels able to model the substitution of a term for a bound variable.
%%%%%%%%%%%%%%%%
\subsection{Contributions}
\paragraph{System $\SBVT$.}
We introduce the system $\SBVT$ (Section~\ref{section:Systems SBVT and BVT})
which extends $\SBV$. The extension of $\SBV$ consists on adding a binary
\emph{renaming} operator $\vlfo{\cdot}{\cdot}$. Renaming is self-dual and binds
atoms. Renaming is the inverse of $\alpha$-rule, its prominent defining axiom
being $\strR \approx \vlfo{\atma}{\strR\subst{\atma}{\atmb}}$. The
meta-operation $\subst{\atma}{\atmb}$ must be a capture-free substitution of the
\emph{atom} $\atma$ for every free occurrence of the atom $\atmb$ in $\strR$ and
of $\natma$ for $\natmb$. The idea is that we shall rename input/output
channels, \ie\ atoms, in formulas that represent  \llcxlinlamterms\
\llcxwithexs. Renaming essentially sets the boundary where the name change can
take place, without altering the set of free names of $\SBVT$ structures.
%%%%%%%%%%%%%%%%%%%%
\paragraph{Completeness of $\SBVT$.}
We define how to transform any \llcxlinlamterm\ \llcxwithexs\ $\llcxM$ into a
formula of $\SBVT$ (Section~\ref{section:Completeness of SBVT and BVT}). Then,
the evaluation of $\llcxM$ becomes a proof-search process inside $\SBVT$:
\begin{quote}
(Theorem~\ref{theorem:Completeness of SBVT},
page~\pageref{theorem:Completeness of SBVT},
Section~\ref{section:Completeness of SBVT and BVT})
For every \llcxlinlamterm\ \llcxwithexs\ $\llcxM$, and every atom $\atmo$, which
plays the role of output-channel, if $\llcxM$ reduces to $\llcxN$, then:
$$\vlderivation                            {
\vlde{}{\SBVT}{\mapLcToDi{\llcxM}{\atmo}}{
\vlhy         {\mapLcToDi{\llcxN}{\atmo}}}}$$
is a derivation of $\SBVT$, with $\mapLcToDi{\llcxM}{\atmo}$ as conclusion, and
$\mapLcToDi{\llcxN}{\atmo}$ as premise.
\end{quote}
Thanks to the deep application of rules, proof-search inside $\SBVT$ is
completely flexible, so it can simulate any evaluation strategy from $\llcxM$ to
$\llcxN$.
%%%%%%%%%%%%%%%%
\paragraph{Completeness of $\BVT$.}
In fact, we can also show that a computation from $\llcxM$ to $\llcxN$ in
\llcxlinlamcalc\ \llcxwithexs\ becomes a process of annihilation between the
formula that represents $\llcxM$, and the \emph{negation} of the formula
representing $\llcxN$:
\begin{quote}
(Corollary~\ref{corollary:Completeness of BVT},
page~\pageref{corollary:Completeness of BVT},
Section~\ref{section:Completeness of SBVT and BVT}.)
For every $\llcxM$, and  $\atmo$, if $\llcxM$ reduces to $\llcxN$, then
$\vlsbr[\mapLcToDi{\llcxM}{\atmo} ;\vlne{\mapLcToDi{\llcxN}{\atmo}}]$ is a
theorem of $\BVT$.
\end{quote}
%%%%%%%%%%%%%%%%
\paragraph{Cut elimination for $\SBVT$.}
The completeness of $\BVT$ follows from proving that the cut elimination holds
inside $\SBVT$ (Theorem~\ref{theorem:Admissibility of the up fragment},
page~\pageref{theorem:Admissibility of the up fragment},
Section~\ref{section:Splitting theorem of BVT}). The proof of cut elimination
extends to $\SBVT$ the path followed to prove the cut-elimination for $\SBV$
\cite{Gugl:06:A-System:kl}, based on the four main steps \emph{shallow
splitting}, \emph{context reduction}, \emph{splitting}, and
\emph{admissibility of the up fragment}.
%%%%%%%%%%%%%%%%%
\paragraph{Soundness of $\SBVT$.}
We show that proof-search of $\SBVT$ can be an interpreter of \llcxlamterms\
\llcxwithexs. Then, the evaluation of $\llcxM$ becomes a proof-search process
inside $\SBVT$:
\begin{quote}
(Theorem~\ref{theorem:Weak soundness of SBVT},
page~\pageref{theorem:Weak soundness  of SBVT},
Section~\ref{section:Completeness of SBVT and BVT}.)
For every \llcxlinlamterm\ \llcxwithexs\ $\llcxM$, and every atom $\atmo$, which
plays the role of output-channel, if
$$\vlderivation                           {
\vlde{}{\SBVT}{\mapLcToDi{\llcxM}{\atmo}}{
\vlhy         {\mapLcToDi{\llcxN}{\atmo}}}}$$
is a derivation of $\SBVT$, with $\mapLcToDi{\llcxM}{\atmo}$ as conclusion, and
$\mapLcToDi{\llcxN}{\atmo}$ as premise, then $\llcxM$ reduces to $\llcxN$.
\end{quote}
%%%%%%%%%%%%%%%%%%
\paragraph{Soundness of $\BVT$.}
We show that proof-search of $\BVT$ can be an interpreter of \llcxlamterms\
\llcxwithexs\ (Section~\ref{section:Completeness of SBVT and BVT}).
\begin{quote}
(Corollary~\ref{corollary:Weak soundness of BVT},
page~\pageref{corollary:Weak soundness of BVT},
Section~\ref{section:Completeness of SBVT and BVT})
For every \llcxlinlamterms\ \llcxwithexs\ $\llcxM, \llcxN$, and every atom
$\atmo$, which plays the role of output-channel, if
$\vlsbr[\mapLcToDi{\llcxM}{\atmo};\vlne{\mapLcToDi{\llcxN}{\atmo}}]$ is a
theorem of $\BVT$, then $\llcxM$ reduces to
$\llcxN$.
\end{quote}
\noindent
In principle, this means that if we think $\llcxM$ reduces to $\llcxN$, we can
check our conjecture by looking a proof of $\vlsbr[\mapLcToDi{\llcxM}{\atmo}
;\vlne{\mapLcToDi{\llcxN}{\atmo}}]$ inside $\BVT$. However, it is worth
remarking that we can prove
$\vlsbr[\mapLcToDi{\llcxM}{\atmo};\vlne{\mapLcToDi{\llcxN}{\atmo}}]$ is a
theorem of $\BVT$ only under a specific proof-search strategy. This, might limit
efficiency. Indeed, the freedom we could gain, at least in principle, thanks to
the deep application of the logical rules, in the course of a proof-search might
be lost by sticking to the specific strategy we are referring to and that we
shall see.
%%%%%%%%%%%%%%%%
\paragraph{Expressiveness of $\SBVT$ and $\BVT$.}
\llcxlinlamcalc\ is $\PTIMEC$, using polynomial time Turing machines as
complexity model of reference \cite{Mairson:2003JFP}. The proof in
\cite{Mairson:2003JFP} shows that \llcxlamcalc\ computes all boolean functions.
So, proof-search of $\SBVT$ and $\BVT$ can do the same. The extension of
$\SBV$ to $\SBVT$ is not trivial.
%%%%%%%%%%%%%%%%
\paragraph{Acknowledgments.}
We like to thanks Paola Bruscoli and Alessio Guglielmi for stimulating
questions and comments that helped improving the presentation of this work.

\section{Systems $\SBVT$ and $\BVT$}
\label{section:Systems SBVT and BVT}

\paragraph{Structures.}
Let $\atma, \atmb, \atmc, \ldots$ denote the elements of a countable set of
\dfn{positive propositional variables}, while $\natma, \natmb, \natmc, \ldots$
denote the set of \dfn{negative propositional variables}, isomorphic to the set
of positive ones. The set of \dfn{atoms} contains positive and negative
propositional variables, and nothing else. Let $\vlone$ be a \dfn{constant}
different from any atom.
%%%%%%%%%
\begin{figure}[ht]
\begin{center}
        \fbox{
                \begin{minipage}{.9\textwidth}
                        %%%%%%%%%%%%%
                        \input{BV2-structures}
                        %%%%%%%%%%%%%
               \end{minipage}
        }%\fbox
\end{center}
\caption{Structures}
\label{fig:BVT-structures}
\end{figure}
%%%%%%%%%
The grammar in Figure~\ref{fig:BVT-structures} gives the set of
\dfn{structures}. The structures \dfn{par}, \dfn{copar}, and \dfn{seq} come from
$\SBV$. \dfn{Renaming} $\vlfo{\atma}{\strR}$ is new and comes with the proviso
that $\atma$ must be a positive atom. Namely, $\vlfo{\natma}{\strR}$ is not in
the syntax. Renaming implies the definition of the \dfn{free names}
$\strFN{\strR}$ of $\strR$ as in Figure~\ref{fig:BVT-free names and
substitution}.
%%%%%%%%%
\begin{figure}[ht]
\begin{tabular}{cc}
  \fbox{
        \begin{minipage}{.95\textwidth}
                %%%%%%%%%%%%%
                \input{BV2-free-names}
                %%%%%%%%%%%%%
       \end{minipage}
  }% \fbox
\end{tabular}
\caption{Free names of structures.}
\label{fig:BVT-free names and substitution}
\end{figure}
%%%%%%%%%%%%%
\paragraph{Size of the structures.}
The \dfn{size} $\Size{\strR}$ of $\strR$ sums the number of occurrences
of atoms in $\strR$ and the number of renaming $\vlfo{\atma}{\strT}$ inside
$\strR$ whose bound variable $\atma$ belongs to $\strFN{\strT}$.
For example, $\vlstore{\vlsbr[\atmb;\natmb]}\Size{\vlfo{\atma}{\vlread}} = 2$,
while $\vlstore{\vlsbr[\atma;\natma]}\Size{\vlfo{\atma}{\vlread}} = 3$.
%%%%%%%%%
\paragraph{Equivalence on structures.}
%%%%%%%%%
\begin{figure}[ht]
\begin{center}
        \fbox{
                \begin{minipage}{.9\textwidth}
                %%%%%%%%%%%%%
                  \begin{center}
%     \vspace{-1.5\baselineskip}
    \begin{tabular}{cc}
      \begin{minipage}{.45\textwidth}
       \begin{center}
        \smalltitle{\textbf{Negation}}%
        {%do not erase
         \begin{eqnarray*}
           \nonumber
           \vlne{\vlone} &
           \approx & \vlone
           \\
           \nonumber
           \vlstore{\vlsbr[\strR;\strT]}
           \vlne{\vlread} &
           \approx & {\vlsbr(\vlne{\strR};\vlne{\strT})}
           \\
           \nonumber
           \vlstore{\vlsbr(\strR;\strT)}
           \vlne{\vlread} &
           \approx & {\vlsbr[\vlne{\strR};\vlne{\strT}]}
           \\
           \nonumber
           \vlstore{\vlsbr<\strR;\strT>}
           \vlne{\vlread} &
           \approx & {\vlsbr<\vlne{\strR};\vlne{\strT}>}
           \\
           \nonumber
           \vlne{\vlfo{\atma}{\strR}} & \approx &
         \end{eqnarray*}
        }%do not erase

        \smalltitle{\textbf{Binder}}%
        {%do not erase
                \begin{eqnarray*}
                  \strR & \approx & \vlfo{\atma}{\strR\subst{\atma}{\atmb}}
                                    \textrm{ if }
                                    \atma\not\in\strFN{\strR}\\
                  \atmb\subst{\atma}{\atmb}  & \approx & \atma          \\
                  \natmb\subst{\atma}{\atmb} & \approx & \natma         \\
                  \atmc\subst{\atma}{\atmb}  & \approx & \atmc          \\
                  \natmc\subst{\atma}{\atmb} & \approx & \natmc         \\
                  {\vlsbr[\strR;\strT]}\subst{\atma}{\atmb} &
                        \approx &
                     {\vlsbr[\strR\subst{\atma}{\atmb};
                             \strT\subst{\atma}{\atmb}]} \\
                  {\vlsbr(\strR;\strT)}\subst{\atma}{\atmb} &
                        \approx &
                     {\vlsbr(\strR\subst{\atma}{\atmb};
                             \strT\subst{\atma}{\atmb})} \\
                  {\vlsbr<\strR;\strT>}\subst{\atma}{\atmb} &
                        \approx &
                     {\vlsbr<\strR\subst{\atma}{\atmb};
                             \strT\subst{\atma}{\atmb}>} \\
                  \vlfo{\atmb}{\strR}\subst{\atma}{\atmb} & \approx & \strR \\
                  \vlfo{\atmc}{\strR}\subst{\atma}{\atmb} & \approx &
                       \vlfo{\atmc}{\strR\subst{\atma}{\atmb}}
                \end{eqnarray*}
        }%do not erase

       \end{center}
      \end{minipage}
      &
      \begin{minipage}{.45\textwidth}
       \begin{center}
%         \vspace{2\baselineskip}
        %%%

        %%%
        \smalltitle{\textbf{Contextual Closure}}\\%
        {%do not erase
        \vspace{\baselineskip}
          ${\strS{\vlscn{\strR}}} \ \approx \ {\strS{\vlscn{\strT}}}$ \ if \
          $\strR \ \approx \ \strT$
        \vspace{\baselineskip}
        }%do not erase

        \smalltitle{\textbf{Unit}}%
        {%do not erase
                \begin{eqnarray*}
%                   \strR           & \approx{\vlsbr[\strR]}
%                                     \approx{\vlsbr(\strR)}
%                                     \approx{\vlsbr<\strR>}
%                   \\
                  \strR  & \approx & \vlsbr[\vlone;\strR] \\
                  \strR  & \approx & \vlsbr<\vlone;\strR>
                         \ \approx \ \vlsbr<\strR;\vlone>
%                   \\
%                   \vlone          & \approx{\vlfo{\atma}{\vlone}}
                \end{eqnarray*}
        }%do not erase

        \smalltitle{\textbf{Associativity}}%
        {%do not erase
          \begin{eqnarray*}
           \vlsbr[[\strR;\strT];\strV]
           & \approx &
           \vlsbr[\strR;[\strT;\strV]]
           \\
           \vlsbr<<\strR;\strT>;\strV>
           & \approx &
           \vlsbr<\strR;<\strT;\strV>>
        \end{eqnarray*}
        }%do not erase

        \smalltitle{\textbf{Commutativity}}%
        {%do not erase
                \begin{eqnarray*}
                 \vlsbr[\strR;\strT]  & \approx & \vlsbr[\strT;\strR]
                \end{eqnarray*}
        }%do not erase

        \smalltitle{\textbf{Distributivity}}%
        {%do not erase
                \begin{eqnarray*}
		  \vlstore{\vlsbr<\strR;\strU>}
		  \vlfo{\atma}{\vlread}
                  & \approx &
		  \vlsbr<{\vlfo{\atma}{\strR}};{\vlfo{\atma}{ \strU}}>
                \end{eqnarray*}
        }%do not erase

        %%%
        \smalltitle{\textbf{Exchange}}%
        {%do not erase
                \begin{eqnarray*}
                 \vlfo{\atma}{\vlex{\atmb}{\strR}}
                  & \approx &
                 \vlex{\atmb}{\vlfo{\atma}{\strR}}
                \end{eqnarray*}
        }%do not erase

       \end{center}
      \end{minipage}
    \end{tabular}
%    \vspace{\baselineskip}
  \end{center}

                %%%%%%%%%%%%%
               \end{minipage}
        }%\fbox
\end{center}
\caption{Equivalence $\approx$ on structures.}
\label{fig:BVT-structure equivalence}
\end{figure}
%%%%%%%%
Structures are equivalent up to the smallest congruence defined by the set of
axioms in Figure~\ref{fig:BVT-structure equivalence} that assigns to renaming
the status of self-dual operator. The reason is intuitive. By definition,
$\strR\subst{\atma}{\atmb}$ substitutes every (free) occurrence of the atom
$\atma$, and its dual $\natma$, for $\atmb$, and $\natmb$, respectively, in
$\strR$. Nothing changes when acting on $\vlne{\strR}$ where every occurrence of
$\atma$ corresponds to one of $\natma$ in $\strR$, and everyone of $\natma$ to
one of $\atma$. Moreover, thanks to negation axioms in
Figure~\ref{fig:BVT-structure equivalence}, the following set of equivalence
axioms holds as well:
$\strR  \approx{\vlsbr(\vlone;\strR)}$,
${\vlsbr((\strR;\strT);\strV)} \approx{\vlsbr(\strR;(\strT;\strV))}$,
${\vlsbr(\strR;\strT)}         \approx{\vlsbr(\strT;\strR)} $,
and, remarkably, both
\dfn{renaming elimination} which says that $\vlex{\atma}{\strR} \equiv \strR$ if
$\atma\not\in\strFN{\strR}$, and \dfn{renaming unit}
$\vlone\approx{\vlfo{\atma}{\vlone}}$.

\paragraph{(Structure) Contexts.}
They are $\strS\vlhole$, \ie\ a structure with a single hole $\vlhole$ in it. If
$\strS\vlscn{\strR}$, then $\strR$ is a \dfn{substructure} of $\strS$. For
example, we shall tend to shorten
$\vlstore{\vlsbr[\strR;\strU]}\strS\vlscn{\vlread}$ as
$\strS{\vlsbr[\strR;\strU]}$ when ${\vlsbr[\strR;\strU]}$ fills the hole
$\vlhole$ of $\strS\vlhole$ exactly.

%%%%%%%%%%%
\begin{figure}[ht]
\begin{center}
\fbox{
        \begin{minipage}{.9\textwidth}
        \begin{center}
        %%%%%%%%%%%%%
        {\small
\begin{tabular}{ccc}
$\vlinf{\bvtatidrule}{}{\vlsbr[\atma;\natma]}
                        {\vlone}$
&
&
$\vlinf{\bvtatiurule}{}{\vlone}
                       {\vlsbr(\atma;\natma)}$
\\
\\
$\vlinf{\bvtseqdrule}{}
        {\vlsbr[<\strR;\strT>;<\strU;\strV>]}
        {\vlsbr<[\strR;\strU];[\strT;\strV]>}$
&\qquad
$\vlinf{\bvtswirule}{}
        {\vlsbr[(\strR;\strU);\strT]}
        {\vlsbr([\strR;\strT];\strU)}$
&\qquad
$\vlinf{\bvtsequrule}{}
        {\vlsbr<(\strR;\strU);(\strT;\strV)>}
        {\vlsbr(<\strR;\strT>;<\strU;\strV>)}$
\\
\\
$\vlinf{\bvtrdrule}{}
       {\vlsbr[{\vlfo{\atma}{\strR}};{\vlex{\atma}{\strU}}]}
       {\vlfo{\atma}{\vlsbr[\strR;\strU]}}$
&\quad&
$\vlinf{\bvtrurule}{}
       {\vlex{\atma}{\vlsbr(\strR;\strU)}}
       {\vlsbr({\vlex{\atma}{\strR}};{\vlfo{\atma}{\strU}})}$
\end{tabular}
}
        %%%%%%%%%%%%%
        \end{center}
        \end{minipage}
    }%\fbox
\end{center}
\caption{System $\SBVT$.}
    \label{fig:SBVT}
\end{figure}

\paragraph{The system $\SBVT$.} 
It contains the set of inference rules in Figure \ref{fig:SBVT} with form
$\vlinf{\bvtrhorule}{}{\strR}{\strT}$, \dfn{name} $\bvtrhorule$, \dfn{premise}
$\strT$, and \dfn{conclusion} $\strR$. One between $\strR$ or $\strT$ may be
missing, but not both. The typical use of an inference rules is
$\vlinf{\bvtrhorule}{}
       {\strS\vlscn{\strR}}
       {\strS\vlscn{\strT}}$. It specifies that if a structure $\strU$ matches
$\strR$ in a context $\strS\vlhole$, it can be rewritten to
$\strS\vlscn{\strT}$. Since rules apply in any context, and we use as rewriting
rules $\strR$ is the \dfn{redex} of $\bvtrhorule$, and $\strT$ its \dfn{reduct}.
\par
The \dfn{down fragment} of $\SBVT$ is $\Set{\vlrulename{\bvtatidrule},
\vlrulename{\bvtswirule}, \vlrulename{\bvtseqdrule}, \vlrulename{\bvtrdrule}}$.
Its \dfn{up fragment} of $\SBVT$ is $\Set{\vlrulename{\bvtatiurule},
\vlrulename{\bvtswirule}, \vlrulename{\bvtsequrule}, \vlrulename{\bvtrurule}}$.
So $\vlrulename{\bvtswirule}$ belongs to both.
\par
Renaming is modeled by $\vlrulename{\bvtrdrule}$, and $\vlrulename{\bvtrurule}$.
The former can be viewed as the restriction to a self-dual quantifier of the
rule $\vlrulename{\mathrm{u}\downarrow}$ which, in \cite{Strasburger:2008qf},
models the universal quantifier.
%%%
\paragraph{Derivation and proof.}
A \dfn{derivation} in $\SBVT$ is either a structure or an instance of the above
rules or a sequence of two derivations. The topmost structure in a derivation is
its \dfn{premise}. The bottommost is its \dfn{conclusion}. The \dfn{size}
$\Size{\bvtDder}$ of a derivation $\bvtDder$ is the number of rule instances in
$\bvtDder$. A derivation $\bvtDder$ of a structure $\strR$ in $\SBVT$ from a
structure $\strT$ in $\SBVT$, only using a subset $\SBVsub\subseteq\SBVT$ is
$\vlderivation                  {
\vlde{\bvtDder}{\SBVsub}{\strR}{
\vlhy                   {\strT}}}
$.
The equivalent \emph{space-saving} form we shall tend to use is
$\bvtInfer{\bvtDder}{\strT \bvtJudGen{\SBVT}{\SBVsub}\strR}$. In general, we
shall drop both $\SBVsub$ and $\SBVT$ when $\bvtDder$ develops in full $\SBVT$.
The derivation
$\vlderivation                  {
\vlde{\bvtDder}{\SBVsub}{\strR}{
\vlhy                   {\strT}}}$
is a \dfn{proof} whenever $\strT\approx \vlone$. We denote it as
$\vlproof{\bvtDder}{\SBVsub}{\strR}$,
$\vlderivation                  {
\vlde{\bvtDder}{\SBVsub}{\strR}{
\vlhy                   {\vlone}}}$,
or $\bvtInfer{\bvtDder}{\
\bvtJudGen{\SBVT}{\SBVsub}\strR}$.
When developing a derivation, we write $\vliqf{}{}{\strR}{\strT}$ to mean
$\strR\approx \strT$. Finally, we shall write
$\vliqf{\Set{\bvtrhorule_1,\ldots,\bvtrhorule_m}}{}{\strR}{\strT}$ if, together
with some equivalence, we apply the set of rules
$\Set{\bvtrhorule_1,\ldots,\bvtrhorule_m}$ to derive $\strR$ from $\strT$.
%%%%%%%%%%
\par
The following proposition shows when two structures $\strR, \strT$ of $\BVT$
can be ``moved'' inside a context so that they are one aside the other and may
eventually communicate going upward in a derivation.
%%%%%%%%%
\begin{proposition}[\textit{\textbf{Context extrusion}}]
\label{proposition:Context extrusion}
$\vlstore{\vlsbr[\strR;\strT]}\strS\vlread\bvtJud{\Set{\vlrulename{\bvtseqdrule}
, \vlrulename{\bvtswirule}, \vlrulename{\bvtrdrule}}}\vlsbr[\strS\vlscn{ \strR }
;\strT]$, for every $\strS, \strR, \strT$.
\end{proposition}
\begin{proof}
By induction on
$\Size{\strS\vlhole}$, proceeding by cases on the form of
$\strS\vlhole$. 
(Details in Appendix~\ref{section:Proof of proposition:Context extrusion}).
\end{proof}
Proposition~\ref{proposition:Context extrusion} here above, also shows how
crucial it is saying that every structure is a derivation of $\BVT$. Otherwise,
the statement would become meaningless in the base case.
%%%%
\paragraph{Equivalence of systems.}
A subset $\SBVsub\subseteq\SBVT$ \dfn{proves} $\strT$ if
$\vlproof{\bvtDder}{\SBVsub}{\strR}$ for some $\bvtDder$. Two subsets $\SBVsub$
and $\SBVsub'$ of the rules in $\SBVT$ are \dfn{strongly equivalent} if, for
every derivation 
$\vlderivation                  {
\vlde{\bvtDder}{\SBVsub}{\strR}{
\vlhy                   {\strT}}}
$, there exists a derivation 
$\vlderivation                    {
\vlde{\bvtDder'}{\SBVsub'}{\strR}{
\vlhy                     {\strT}}}
$, and vice versa.
Two \dfn{systems are equivalent} if they prove the same structures.
%%%%%%%%%
\paragraph{Admissible and derivable rules.}
A rule $\bvtrhorule$ is \dfn{admissible} for the system $\SBVT$ if
$\bvtrhorule\notin\SBVT$ and, for every derivation $\bvtDder$ \ST\
$\vlderivation                                     {
\vlde{\bvtDder}{\Set{\bvtrhorule}\cup\SBVT}{\strR}{
\vlhy                                      {\strT}}}
$, there is a derivation $\bvtDder'$ \ST\
$\vlderivation                 {
\vlde{\bvtDder'}{\SBVT}{\strR}{
\vlhy                  {\strT}}}
$. A rule $\bvtrhorule$ is \dfn{derivable} in $\SBVsub\subseteq\SBVT$ if
$\bvtrhorule\not\in\SBVsub$ and, for every instance
$\vlinf{\bvtrhorule}{}{\strR}{\strT}$ there exists a derivation $\bvtDder$ in
$\SBVsub$ \ST\ $\vlder{\bvtDder}{\SBVsub}{\strR}{\strT}$.
%%%%%%%%%%%%
\begin{figure}[ht]
\begin{center}
        \fbox{
        \begin{minipage}{.9\textwidth}
        \begin{center}
        %%%%%%%%%%%%%
        {\small
\begin{tabular}{ccccc}
$\vlinf{\bvtintdrule}{}{\vlsbr[\strR;\nstrR]}
                       {\vlone}$                                       &\quad&
$\vlinf{\bvtinturule}{}{\vlone}
                       {\vlsbr(\strR;\nstrR)}$                       &\quad&
$\vlinf{\bvtmixprule}{}{\vlsbr<\strR;\strT>}
                       {\vlsbr(\strR;\strT)}$ \\\\
$\vlinf{\bvtdefdrule}{}{\vlsbr[<\strR;\natma>;(\atma;\strT)]}
                       {\vlsbr<\strR;\strT>}$                        &\quad&
$\vlinf{\bvtdefurule}{}{\vlsbr<\strR;\strT>}
                       {\vlsbr(<\strR;\natma>;[\atma;\strT])}$       &\quad&
$\vlinf{\bvtpmixrule}{}{\vlsbr[\strR;\strT]}
                       {\vlsbr<\strR;\strT>}$
\end{tabular} 
}
        %%%%%%%%%%%%%
        \end{center}
        \end{minipage}
        }%\fbox
\end{center}
\caption{A core-set of rules derivable in $\SBVT$.}
\label{fig:SBVT-core-set-derivable-rules}
\end{figure}
%%%%%%%%%%%
Figure~\ref{fig:SBVT-core-set-derivable-rules} shows a core set of rules
derivable in $\SBVT$. The rules $\vlrulename{\bvtintdrule}$, and
$\vlrulename{\bvtinturule}$ are the \dfn{general interaction down} and \dfn{up},
respectively. The rule $\vlrulename{\bvtdefdrule}$ uses $\natma$ as a
place-holder and $\atma$ as name for $\strT$. Building the derivation upward, we
literally replace $\strT$ for $\natma$. Symmetrically for
$\vlrulename{\bvtdefurule}$. The rules $\vlrulename{\bvtmixprule}$, and
$\vlrulename{\bvtpmixrule}$ show a hierarchy between the connectives, where
$\vlpa$ is the lowermost, $\vlse$ lies in the middle, and $\vlte$ on top.
%%%
\paragraph{\dfn{General interaction up} is derivable in
$\Set{\vlrulename{\bvtinturule}, \vlrulename{\bvtswirule},
\vlrulename{\bvtsequrule}, \vlrulename{\bvtrurule}}$.} We can prove it by
induction on $\Size{\strR}$, proceeding by cases on the form of $\strR$. We
detail out the only the case new to $\BVT$. Let $\strR \equiv
{\vlfo{\atma}{\strT}}$. Then:
\[
\vlderivation                                                             {
\vliq{            }{}{\vlone}                                            {
\vlin{\bvtinturule}{}{\vlex{\atma}{\vlone}}                             {
\vlin{\bvtrurule}  {}{\vlex{\atma}{\vlsbr(\strT;\nstrT)}}              {
\vlhy                {\vlsbr(\vlfo{\atma}{\strT};\vlex{\atma}{\nstrT})}}}}}
\]
Symmetrically, \dfn{general interaction down} is derivable in
$\Set{\vlrulename{\bvtintdrule}, \vlrulename{\bvtswirule},
\vlrulename{\bvtseqdrule}, \vlrulename{\bvtrdrule}}$.
%%%
\paragraph{The rule $\vlrulename{\bvtdefdrule}$ is derivable in
$\Set{\vlrulename{\bvtatidrule}, \vlrulename{\bvtswirule},
\vlrulename{\bvtseqdrule}}$} as follows:
\[
\vlderivation                                                         {
\vliq{            }{}{\vlsbr[<\strR;\atma >;(\natma;\strT)]         } {
\vlin{\bvtseqdrule}{}{\vlsbr[<\strR;\atma >;<\vlone;(\natma;\strT)>]} {
\vliq{\bvtswirule }{}{\vlsbr<[\strR;\vlone];[\atma ;(\natma;\strT)]>} {
\vlin{\bvtatidrule}{}{\vlsbr<\strR;([\atma;\natma]; \strT)>         } {
\vliq{            }{}{\vlsbr<\strR;(\vlone; \strT)>                 } {
\vlhy                {\vlsbr<\strR;\strT>                       }}}}}}}
\]
Symmetrically, $\vlrulename{\bvtdefurule}$ is derivable in
$\Set{\vlrulename{\bvtatiurule}, \vlrulename{\bvtswirule},
\vlrulename{\bvtsequrule}}$.
%%%%
\paragraph{The rule $\vlrulename{\bvtmixprule}$ is derivable in
$\Set{\bvtsequrule}$} as
follows:
\[
\vlderivation                                                         {
\vliq{            }{}{\vlsbr<\strR;\strT>                           } {
\vlin{\bvtsequrule}{}{\vlsbr<(\strR;\vlone);(\vlone;\strT)>         } {
\vliq{            }{}{\vlsbr(<\strR;\vlone>;<\vlone;\strT>)         } {
\vlhy                {\vlsbr(\strR;\strT)                           }}}}}
\]
Symmetrically, $\vlrulename{\bvtpmixrule}$ is derivable in
$\Set{\vlrulename{\bvtseqdrule}}$.
\section{Splitting theorem of $\BVT$}
\label{section:Splitting theorem of BVT}
The goal is to prove that $\SBVT$, and $\BVT$ are strongly equivalent. Namely,
if a derivation of $\strT$ from $\strR$ exists in one of the two systems, then
there is a derivation of $\strT$ from $\strR$ into the other. Proving the
equivalence, amounts to proving that every up rule is admissible in $\BVT$ or,
equivalently, that we can eliminate them from any derivation of $\SBVT$.
Splitting theorem for $\BVT$, which extends the namesake theorem for $\BV$
\cite{Gugl:06:A-System:kl}, is the effective tool we prove to exist to show
that the up fragment of $\SBVT$ is admissible for $\BVT$.
%%%%%%%%%%%
\begin{proposition}[\textit{\textbf{$\BVT$ is affine}}]
\label{proposition:BVT is affine}
In every derivation $\bvtInfer{\bvtDder} {\strT \bvtJud{} \strR}$, we have
$\Size{\strR} \geq \Size{\strT}$.
\end{proposition}
%%%%%%
\begin{proof}
By induction on $\Size{\bvtDder}$, proceeding by cases on its last rule
$\bvtrhorule$.
\end{proof}

\begin{proposition}[\textit{\textbf{Derivability of structures in $\BVT$}}]
\label{proposition:Derivability of structures in BVT}
For all structures $\strR, \strT$:
%%%%%%%%%
\begin{enumerate}

\item\label{enum:Derivability of subformulas-seq}
$\vlstore{\vlsbr<\strR;\strT>}
 \bvtInfer{\bvtDder}
          {\ \bvtJud{} \vlread}$
\IFF\
$\bvtInfer{\bvtDder_1}
          {\ \bvtJud{} \strR}$ and
$\bvtInfer{\bvtDder_2}{\ \bvtJud{} \strT}$.

\item\label{enum:Derivability of subformulas-copar}
$\vlstore{\vlsbr(\strR;\strT)}
 \bvtInfer{\bvtDder}{\ \bvtJud{} \vlread}$
\IFF\
$\bvtInfer{\bvtDder_1}{\ \bvtJud{} \strR$ and
$\bvtInfer{\bvtDder_2}{\ \bvtJud{} \strT}}$.

\item\label{enum:Derivability of subformulas-fo}
$\bvtInfer{\bvtDder}{\ \bvtJud{} \vlfo{\atma}{\strR}}$
\IFF\
$\bvtInfer{\bvtDder'}{\ \bvtJud{} \strR\subst{\atmb}{\atma}}$,
for every atom $\atmb$.
\end{enumerate}
\end{proposition}
%%%%%%
\begin{proof}
Both \ref{enum:Derivability of subformulas-seq} and \ref{enum:Derivability of
subformulas-copar} hold in $\BV$ \cite{Gugl:06:A-System:kl} while, of course,
\ref{enum:Derivability of subformulas-fo} is meaningless in $\BV$. 
\par
We start proving the \emph{``if implication''}. First, we observe that the
proofs of \ref{enum:Derivability of subformulas-seq} and \ref{enum:Derivability
of subformulas-copar}, given in \cite{Gugl:06:A-System:kl} by induction on
$\Size{\bvtDder}$ inside $\BV$, obviously extend to the cases when the last rule
of $\bvtDder$ is $\vlrulename{\bvtrdrule}$. The reason is that the redex of
$\vlrulename{\bvtrdrule}$ can only be inside $\strR$ or $\strT$.
Concerning~\ref{enum:Derivability of subformulas-fo}, the assumption implies the
existence of $\bvtInfer{\bvtDder'}{\ \bvtJud{} \strR\subst{\atma}{\atma}}$,
namely of $\bvtInfer{\bvtDder'}{\,\bvtJ \strR}$. So, we can ``wrap'' $\bvtDder'$
with $\vlfo{\atma}{\cdot}$ by wrapping the topmost $\vlone$ thanks to $\vlone
\approx {\vlfo{\atma}{\vlone}}$.
\par
For proving the \emph{``only if''} direction we use induction on
$\Size{\bvtDder}$, proceeding by cases on its last rule
$\vlrulename{\bvtrhorule}$. In all the three cases the redex of
$\vlrulename{\bvtrhorule}$ can only be inside $\strR$ or $\strT$. So, the
statements hold by obviously applying the inductive hypotheses.
\end{proof}
%%%%%%%%%%%%%%%5
\begin{proposition}[\textit{\textbf{Shallow Splitting}}]
\label{proposition:Shallow Splitting}
For all structures $\strR, \strT$ and $\strP$:
\begin{enumerate}
\item\label{enum:Shallow-Splitting-atom}
If $\vlstore{\vlsbr[\atma;\strP]}
    \bvtInfer{\bvtDder}
             {\ \bvtJ
                {\vlread}}$, then
$\bvtInfer{\bvtDder'} {\natma \bvtJud{} \strP}$.

\item\label{enum:Shallow-Splitting-seq}
If $\vlstore{\vlsbr[<\strR;\strT>;\strP]}
    \bvtInfer{\bvtDder}
             {\ \bvtJud{} \vlread}$, then
${\vlsbr<\strP_1;\strP_2>} \bvtJud{} \strP$,
$ \bvtJud{} {\vlsbr[\strR;\strP_1]}$, and
$ \bvtJud{} {\vlsbr[\strT;\strP_2]}$, for some
$\strP_1, \strP_2$.

 \item\label{enum:Shallow-Splitting-copar}
If $\vlstore{\vlsbr[(\strR;\strT);\strP]}
     \bvtInfer{\bvtDder}{\ \bvtJud{} \vlread}$,
then
${\vlsbr[\strP_1;\strP_2]} \bvtJud{} \strP$,
$ \bvtJud{} {\vlsbr[\strR;\strP_1]}$, and
$ \bvtJud{} {\vlsbr[\strT;\strP_2]}$, for some
$\strP_1, \strP_2$.

 \item\label{enum:Shallow-Splitting-fo}
If $\vlstore{\vlsbr[\vlfo{\atma}{\strR};\strP]}
    \bvtInfer{\bvtDder} {\ \bvtJud{} \vlread}$, then
$\vlex{\atma}{\strP'} \bvtJud{} \strP$, and
$ \bvtJud{} {\vlsbr[\strR;\strP']}$, for some $\strP'$.

\end{enumerate}
\end{proposition}
%%%
\begin{proof}
Point~\ref{enum:Shallow-Splitting-atom} holds by induction on $\Size{\bvtDder}$,
reasoning by cases on the last rule $\bvtrhorule$ of $\bvtDder$.
\par
From \cite{Gugl:06:A-System:kl} we know that the statements
\ref{enum:Shallow-Splitting-seq} and \ref{enum:Shallow-Splitting-copar} hold in
$\BV$ by induction on the lexicographic order of the pair $(\Size{\strV},
\Size{\bvtDder})$, where $\strV$ is one between ${\vlsbr[<\strR;\strT>;\strP]}$
or ${\vlsbr[(\strR;\strT);\strP]}$, proceeding by cases on the last rule
$\bvtrhorule$ of $\bvtDder$. The proof of
points~\ref{enum:Shallow-Splitting-seq}, and~\ref{enum:Shallow-Splitting-copar}
extends to the cases where $\bvtrhorule$ is $\vlrulename{\bvtrdrule}$, using the
same inductive measure.
\par
Also point~\ref{enum:Shallow-Splitting-fo} holds by induction on the above
lexicographic order of the pair $(\Size{\strV}, \Size{\bvtDder})$.
(Details in Appendix~\ref{section:Proof of proposition:Shallow Splitting}).
\end{proof}
%%%%%%%%%%%%%%%%%%%%%%%%%
\begin{proposition}[\textit{\textbf{Context Reduction}}]
\label{proposition:Context Reduction}
For all structures $\strR$ and contexts $\strS\vlhole$ \ST\
$\bvtInfer{\bvtDder}{\ \bvtJud{} \strS\vlscn{\strR}}$, there are $\strU, \atma$
\ST\
$\vlstore{\vlsbr[\vlhole;\strU]}
 \bvtInfer{\bvtPder}
          {\vlfo{\atma}{\vlread} \bvtJud{} \strS\vlhole}$,
and $\bvtJud{} {\vlsbr[\strR;\strU]}$.
\end{proposition}
\begin{proof}
The proof is by induction on $\Size{\strS\vlhole}$, proceeding by cases on the
form of $\strS\vlhole$. (Details in Appendix~\ref{section:Proof of
proposition:Context Reduction}).
\end{proof}
\begin{remark}[\textbf{\textit{Reading correctly
Proposition~\ref{proposition:Context Reduction}}}]
The statement here above is a compressed version of the more explicit one here
below:
\begin{quote}
For all structures $\strR$ and contexts $\strS\vlhole$ \ST\
$\bvtInfer{\bvtDder}{\ \bvtJud{} \strS\vlscn{\strR}}$, there are $\strU, \atma$
\ST, for every structure $\strV$, $\vlstore{\vlsbr[\vlscn{\strV};\strU]}
\bvtInfer{\bvtPder} {\vlfo{\atma}{\vlread} \bvtJud{} \strS\vlscn{\strV}}$, and
$\bvtJud{} {\vlsbr[\strR;\strU]}$.
\end{quote}
Namely, $\strS\vlhole$ supplies the ``context'' $\strU$, required for proving 
$\strR$, no matter which structure fills the hole of $\strS\vlhole$.
\end{remark}
%%%%%%%%%%%%%%%%%%%
\begin{theorem}{(\textit{\textbf{Splitting}}.)}
\label{theorem:Splitting}
For all structures $\strR, \strT$ and contexts $\strS\vlhole$:
\begin{enumerate}
 \item\label{enum:Splitting-seq}
If $\vlstore{\vlsbr<\strR;\strT>}
    \bvtInfer{\bvtDder}{\ \bvtJud{} \strS{\vlread}}$,
then
$\vlstore{\vlsbr[\vlhole;<\strK_1;\strK_2>]}
 \vlfo{\atma}{\vlread} \bvtJud{} \strS\vlhole$,
$ \bvtJud{} {\vlsbr[\strR;\strK_1]}$, and
$ \bvtJud{} {\vlsbr[\strT;\strK_2]}$, for some $\strK_1, \strK_2, \atma$.

 \item\label{enum:Splitting-copar}
If $\vlstore{\vlsbr(\strR;\strT)}
    \bvtInfer{\bvtDder}{\ \bvtJud{} \strS{\vlread}}$,
then
$\vlstore{\vlsbr[\vlhole;[\strK_1;\strK_2]]}
 \vlfo{\atma}{\vlread} \bvtJ\strS\vlhole$,
$ \bvtJud{} {\vlsbr[\strR;\strK_1]}$, and
$ \bvtJud{} {\vlsbr[\strT;\strK_2]}$, for some $\strK_1, \strK_2, \atma$.

 \item\label{enum:Splitting-fo-ex}
If $\vlstore{\vlfo{\atma}{\strR}}
    \bvtInfer{\bvtDder}{\ \bvtJud{} \strS{\vlread}}$,
then
$\vlstore{\vlsbr[\vlhole;\strK]}
 \vlfo{\atma}{\vlread} \bvtJud{} \strS\vlhole$, and
$ \bvtJud{} {\vlsbr[\strR;\strK]}$, for some $\strK, \atma$.
\end{enumerate}
\end{theorem}
\begin{proof}
We obtain the proof of the three statements by composing
Context Reduction (Proposition~\ref{proposition:Context Reduction}), and
Shallow Splitting (Proposition~\ref{proposition:Shallow Splitting}) in this
order.
(Details in Appendix~\ref{section:Proof of theorem:Splitting}).
\end{proof}
%%%%%%%%%%%%%%%%%%%%%%%%%%%%%%%%%%%%%55
\section{Cut elimination of $\SBVT$}
\label{section:Cut elimination of SBVT}
%%%%%%%%%%%%%
\begin{theorem}[\textit{\textbf{Admissibility of the up fragment}}]
\label{theorem:Admissibility of the up fragment} The up fragment
$\Set{\vlrulename{\bvtatiurule}, \vlrulename{\bvtsequrule},
\vlrulename{\bvtrurule}}$ of $\SBVT$ is admissible for $\BVT$.
\end{theorem}
\begin{proof}
Using splitting (Theorem~\ref{theorem:Splitting}) and shallow splitting
(Proposition~\ref{proposition:Shallow Splitting}) it is enough to show that: (i)
$\vlrulename{\bvtatiurule}$ gets replaced by a derivation that contains an
instance of $\vlrulename{\bvtatidrule}$,
(ii) $\vlrulename{\bvtsequrule}$ gets replaced by a derivation that contains a
couple of instances of $\vlrulename{\bvtseqdrule}$ and
$\vlrulename{\bvtswirule}$ rules,
(iii) $\vlrulename{\bvtrurule}$ gets replaced by a derivation that contains a
couple of instances of $\vlrulename{\bvtrdrule}$ and $\vlrulename{\bvtswirule}$
rules. (Details in Appendix~\ref{section:Proof of theorem:Admissibility of the
up fragment}).
\end{proof}
%%%%%%%%%%%%%%%
Theorem~\ref{theorem:Admissibility of the up fragment} here above directly
implies:
\begin{corollary}
\label{corollary:Cut elimination}
The cut elimination holds for $\SBVT$.
\end{corollary}
\section{\llcxLinlamcalc\ \llcxwithexs}
It is a pair with a set of \llcxlinlamterms, and an operational semantics on
them. The operational semantics looks at substitution as explicit syntactic
component and not as meta-operation.
%%%%%%%%%%%%%%5
\paragraph{The \llcxlinlamterms.}
Let $\llcxVar$ be a countable set of variable names we range over by
$\llcxX,\llcxY,\llcxW,\llcxZ$. We call $\llcxVar$ the
\dfn{set of \llcxlamvars}.
The set of \dfn{\llcxlinlamterms\ \llcxwithexs} is
$\llcxSet{}  = \bigcup_{X\subset\llcxVar}\llcxSet{X}$ we range over by
$\llcxM,\llcxN,\llcxP,\llcxQ$. For every $X\subset\llcxVar$, the set
$\llcxSet{X}$ contains the
\dfn{\llcxlinlamterms\ \llcxwithexs\ whose free
variables are in $X$}, and which we define as follows:
(i)   $\llcxX\in\llcxSet{\Set{\llcxX}}$;
(ii)  $\llcxF{\llcxX}{\llcxM}\in\llcxSet{X}$
      if $\llcxM\in\llcxSet{X\cup\Set{\llcxX}}$;
(iii) $\llcxA{\llcxM}{\llcxN}\in\llcxSet{X\cup Y}$
      if  $\llcxM\in\llcxSet{X}$,
          $\llcxN\in\llcxSet{Y}$, and
          $X\cap Y=\emptyset$;
(iv)  $\llcxE{\llcxM}{\llcxX}{\llcxP}\in\llcxSet{X\cup Y}$
      if  $\llcxM\in\llcxSet{X\cup\Set{\llcxX}}$,
          $\llcxP\in\llcxSet{Y}$, and
          $X\cap Y=\emptyset$.

\paragraph{$\beta$-reduction on \llcxlinlamterms\ \llcxwithexs .}
\begin{figure}%%%
\begin{center}
\fbox{
\begin{minipage}{.9\textwidth}
%%%%% Alternative useless???? Reduction
{\small
%%%%%%%%%%%%
\input{LLCXS-beta-reduction}
%%%%%%%%%%%%
}
\end{minipage}
}%\fbox
\end{center}
\caption{$\beta$-reduction $\llcxBeta \subseteq\llcxSet{}\times \llcxSet{}$
with
explicit substitution.}
\label{fig:beta-reduction llcxBeta}
\end{figure}
%%%%%%%%%%%%
It is the relation $\llcxBeta$ in Figure~\ref{fig:beta-reduction llcxBeta}. It
is the core of the very simple, indeed, computations the syntax of the terms in
$\llcxSet{}$  allow to develop. The point, however, is that the computational
mechanism that replaces a terms for a variable is there, and we aim at modeling
it inside $\BVT$.
%%%%%%%%%%%%%%5
\paragraph{Operational semantics on \llcxlinlamterms\ \llcxwithexs.}
%%%%
\begin{figure}%%%
\begin{center}
\fbox{
\begin{minipage}{.9\textwidth}
%%%%% Alternative useless???? Reduction
{\small
%%%%%%%%%%%%
\begin{center}
\begin{tabular}{ccccccc}
$\vlinf{\llcxSOSrflrule}{}
       {\llcxSOSJud{\llcxM}{\llcxM}}
       {}
$ &\quad&
$\vlinf{\llcxSOSBrule}{}
       {\llcxSOSJud{\llcxM}{\llcxN}}
       {\llcxM \llcxBeta \llcxN}
$ &\quad&
$\vlinf{\llcxSOStrarule}{}
       {\llcxSOSJud{\llcxM}{\llcxN}}
       {\llcxSOSJud{\llcxM}{\llcxP}
        \qquad
        \llcxSOSJud{\llcxP}{\llcxN}}
$ \\\\
$\vlinf{\llcxSOSfrule}{}
       {\llcxSOSJud{\llcxF{\llcxX}{\llcxM}}{\llcxF{\llcxX}{\llcxN}}}
       {\llcxSOSJud{\llcxM}{\llcxN}}
$  &\quad&
$\vlinf{\llcxSOSalrule}{}
       {\llcxSOSJud{\llcxA{\llcxM}{\llcxP}}{\llcxA{\llcxN}{\llcxP}}}
       {\llcxSOSJud{\llcxM}{\llcxN}}
$ &\quad& 
$\vlinf{\llcxSOSarrule}{}
       {\llcxSOSJud{\llcxA{\llcxP}{\llcxM}}{\llcxA{\llcxP}{\llcxN}}}
       {\llcxSOSJud{\llcxM}{\llcxN}}
$ \\\\
$\vlinf{\llcxSOSslrule}{}
       {\llcxSOSJud{\llcxE{\llcxM}{\llcxX}{\llcxP}}
                   {\llcxE{\llcxN}{\llcxX}{\llcxP}}}
       {\llcxSOSJud{\llcxM}{\llcxN}}
$ &\quad&
  &\quad&
$\vlinf{\llcxSOSsrrule}{}
       {\llcxSOSJud{\llcxE{\llcxP}{\llcxX}{\llcxM}}
                   {\llcxE{\llcxP}{\llcxX}{\llcxN}}}
       {\llcxSOSJud{\llcxM}{\llcxN}}
$ 
\end{tabular}
\end{center}

%%%%%%%%%%%%
}
\end{minipage}
}%\fbox
\end{center}
\caption{Rewriting relation $\llcxSOSJud{}{}\subseteq\llcxSet{}\times
\llcxSet{}$.}
\label{fig:Rewriting relation llcxSOSJud}
\end{figure}
%%%%%%%%%%%%
It is the relation $\,\llcxSOSJud{\!}{}$ in Figure~\ref{fig:Rewriting
relation llcxSOSJud}. It is the contextual and transitive closure of the above
$\beta$-reduction with explicit substitution. We denote as
$\Size{\llcxSOSJud{\llcxM}{\llcxN}}$ the number of instances of rules in
Figure~\ref{fig:Rewriting relation llcxSOSJud}, used to derive
$\llcxSOSJud{\llcxM}{\llcxN}$.
\section{Completeness and Soundness of $\SBVT$ and $\BVT$}
\label{section:Completeness of SBVT and BVT}
We relate functional and proof-theoretic worlds. First we map terms of
$\llcxSet{}$ into structures of $\SBVT$. Then, we show the \emph{completeness}
of $\SBVT$ and $\BVT$, \ie\ that the computations of $\llcxSet{}$ correspond to
proof-search inside the two systems. Finally, we prove \emph{soundness} of
$\SBVT$ and  $\BVT$ \wrt\ the computations of \llcxlamcalc\ \llcxwithexs\ under
a specific proof-search strategy. This means that we can use $\SBVT$ or $\BVT$
to compute any term which any given $\llcxM$ reduces to.
%%%%%%%%%%%%%%%%%%%%%%%%%%%%%
\paragraph{The map $\mapLcToDi{\cdot}{\cdot}$.}
We start with the following ``fake'' map from $\llcxSet{}$ to $\SBVT$:
{\small{\renewcommand{\vlex}[2]{\exists{#1}.{#2}}
 \renewcommand{\vlfo}[2]{\forall{#1}.{#2}}
{\small
\begin{eqnarray}
\label{eqnarray:fake-map-var}
\mapLcToDi{\llcxX}{\atmo} & = &
{\vlsbr<\llcxX;\natmo>}
\\
\label{eqnarray:fake-map-fun}
\vlstore{\vlsmallbrackets\aftrianglefalse
         \vlfo{\llcxX}
              {\vlex{\atmp}{\vlsbr[\mapLcToDi{\llcxM}{\atmp};(\atmp;\natmo)]}}
        }
\mapLcToDi{\llcxF{\llcxX}{\llcxM}}{\atmo} & = & \vlread
\\
\label{eqnarray:fake-map-app}
\vlstore{
 \vlex{\atmp}
     {{\vlsmallbrackets
       \vlsbr[\mapLcToDi{\llcxM}{\atmp}
             ;\vlex{\atmq}{\mapLcToDi{\llcxN}{\atmq}}
             ;(\atmp;\natmo)]}}
        }
\mapLcToDi{\llcxA{\llcxM}{\llcxN}}{\atmo} & = & \vlread
\\
\label{eqnarray:fake-map-subst}
\vlstore{
 \vlfo{\llcxX}
      {{\vlsmallbrackets
        \vlsbr[\mapLcToDi{\llcxM}{\atmo};\mapLcToDi{\llcxP}{\llcxX}]}}
}
\mapLcToDi{\llcxE{\llcxM}{\llcxX}{\llcxP}}{\atmo} & = & \vlread
\end{eqnarray}
}} 
}
\noindent
We use it only to intuitively illustrate how we shall effectively represent
terms of $\llcxSet{}$ as structures of $\SBVT$. The map here above translates
$\llcxM$ into $\mapLcToDi{\llcxM}{\atmo}$ where $\atmo$ is a unique output
channel, while the whole expression depends on a set of free input channels,
each for every free variable of $\llcxM$. Clause~\eqref{eqnarray:fake-map-var}
associates the input channel $\llcxX$ to the fresh output channel $\natmo$,
under the intuition that $\llcxX$ is \emph{forwarded} to $\atmo$, using the
terminology of \cite{HondaYoshida:TCS1995}. Clause~\eqref{eqnarray:fake-map-fun}
assumes $\mapLcToDi{\llcxM}{\atmp}$ has $\atmp$ as output and (at least)
$\llcxX$ as input. It renames $\atmp$, hidden by $\exists$, as $\natmo$ thanks
to $\vlsbr(\atmp;\natmo)$. This must work for every input $\llcxX$. For this
reason we hide $\llcxX$ by means of $\forall$.
Clause~\eqref{eqnarray:fake-map-app} makes the output channels of both
$\mapLcToDi{\llcxM}{\atmp}$ and $\mapLcToDi{\llcxN}{\atmq}$  local, while
renaming $\atmp$ to $\natmo$ thanks to $\vlsbr(\atmp;\natmo)$. If
$\mapLcToDi{\llcxM}{\atmp}$ will result in the translation of a
$\lambda$-abstraction $\llcxF{\llcxZ}{\llcxP}$, then the existential quantifier
immediately preceding $\mapLcToDi{\llcxN}{\atmq}$ will interact with the
universal quantifier in front of $\mapLcToDi{\llcxM}{\atmp}$. The result will be
an on-the-fly channel name renaming. Clause~\eqref{eqnarray:fake-map-subst}
identifies the output of $\mapLcToDi{\llcxP}{\llcxX}$ with one of the existing
free names of $\mapLcToDi{\llcxM}{\llcxo}$. The identification becomes local
thanks to the universal quantifier.
%%%%%%%%%%%%%
\begin{figure}%%%
\begin{center}
\fbox{
\begin{minipage}{.9\textwidth}
{\small
%%%%%%%%%%%%
\input{LLCXS-to-SBV2-map}
%%%%%%%%%%%%
}
\end{minipage}
}%\fbox
\end{center}
\caption{Map $\mapLcToDi{\cdot}{}$ from $\llcxSet{}$ to structures}
\label{fig:Map from llcxSet to structures}
\end{figure}
%%%%%%%%%%
\par
However, in a setting where the second order quantifiers $\forall$, and
$\exists$ only operate on atoms, distinguishing between the two is meaningless.
So, the renaming can be self-dual and the true map
$\mapLcToDi{\cdot}{\cdot}$  which adheres to the above intuition is in
Figure~\ref{fig:Map from llcxSet to structures}.
\par
We keep stressing that $\mapLcToDi{\cdot}{\cdot}$ strongly recalls output-based
embedding of \emph{standard} \llcxlamcalc\ \llcxwithexs\ into $\pi$-calculus
\cite{DBLP:conf/concur/BakelV09}. In principle, this means that extending
$\SBVT$ with the right logical operators able to duplicate atoms, and
consequently upgrading $\mapLcToDi{\cdot}{\cdot}$, we could model full
$\beta$-reduction as proof-search.
%%%%%%%%%%%%%%%%%%%%%%%%%%%%%%%%%%%%%%%%%%%%%
\begin{figure}[ht]
\begin{center}
        \fbox{
                \begin{minipage}{.9\textwidth}
                        %%%%%%%%%%%%%
                        \begin{center}
{\small
\begin{tabular}{cc}
$\vlinf{\mllax}{}
        {\mllA \mllJud\mllA}
        {}$
&\quad
$\vlinf{\mllcut}{}
        {\mllA_1,\ldots,\mllA_m
        ,\mllB_1,\ldots,\mllB_n \mllJud\mllB}
        {\mllA_1,\ldots,\mllA_m \mllJud\mllA
         \quad
         \mllA,\mllB_1,\ldots,\mllB_n \mllJud\mllB
        }$
\\\\
$\vlinf{\mllimpl}{}
        {\mllA_1,\ldots,\mllA_m \mllJud\mllB\mllimpl\mllA}
        {\mllA_1,\ldots,\mllA_m,\mllB \mllJud\mllA}$
&\quad
$\vlinf{\mlltens}{}
        {\mllA_1,\ldots,\mllA_m
        ,\mllB_1,\ldots,\mllB_n \mllJud\mllA\mlltens\mllB}
        {\mllA_1,\ldots,\mllA_m \mllJud\mllA
         \quad
         \mllB_1,\ldots,\mllB_n \mllJud\mllB
        }$
\end{tabular}
}
\end{center}

                        %%%%%%%%%%%%%
               \end{minipage}
        }%\fbox
\end{center}
\caption{Intuitionistic multiplicative fragment $\IMLL$ of linear logic}
\label{fig:MLL-sequent-calculus}
\end{figure}
%%%%%%%%%%%%%%%%%%%%%%%%%%%%%%%%%%%%%%%%%%%%%
\subsection{Origins of the embedding $\mapLcToDi{\cdot}{\cdot}$}
The very source of this work, hence of the map $\mapLcToDi{\cdot}{\cdot}$, have
been:
\begin{enumerate}
\item
\label{enumerate:trivial observation on MLL}
An almost trivial observation on the form of the derivations of the
\dfn{intuitionistic and multiplicative fragment} of \emph{linear logic}
($\IMLL$) \cite{GirardTaylorLafont89}, recalled in
Figure~\ref{fig:MLL-sequent-calculus}.
\item
The internalization of the notion of $\IMLL$ \emph{sequent}, usually a
meta-notion, inside $\SBV$.
\end{enumerate}
\noindent
The formalization of the trivial observation we mention in
point~\eqref{enumerate:trivial observation on MLL} here above is:
%%%%%
\begin{proposition}
Every derivation $\mllPder$ of the sequent $\mllA_1,\ldots,\mllA_m \mllJud\mllA$
starts from, at least, $m$ instances of the rule $\vlrulename{\mllax}$, each
proving the sequent $\mllA_i\mllJud\mllA_i$, for $1\leq i\leq m$. We call
\emph{free axioms} the set of such instances of $\vlrulename{\mllax}$.
\end{proposition}
%%%%%%%%%
\begin{figure}[ht]
\begin{center}
        \fbox{
                \begin{minipage}{.9\textwidth}
                        %%%%%%%%%%%%%
                        {\small
\begin{center}
\begin{tabular}{rl}
$\mapMLLToBVT{\mllVarA}$
=& \!\!\!\!\!\!
$\mllVarA$
\\
$\mapMLLToBVT{(\mllVarA\mllTens\mllVarB)}$
=& \!\!\!\!\!\!
$\vlsbr(\mapMLLToBVT{\mllVarA};\mapMLLToBVT{\mllVarB})$
\\
$\mapMLLToBVT{(\mllVarA\mllImpl\mllVarB)}$
=& \!\!\!\!\!\!
$\vlsbr[\vlne{\mapMLLToBVT{\mllVarA}}
          ;\mapMLLToBVT{\mllVarB}]$
\\\\
$\mapMLLToBVT{(\mllVarA_1,\ldots,\mllVarA_m\mllJud\mllVarB)}$
=& \!\!\!\!\!\!
$\vlsbr<(\vlne{\mapMLLToBVT{\mllVarA_1}}
           ;\vldots;
           \vlne{\mapMLLToBVT{\mllVarA_m}})
         ;\mapMLLToBVT{\mllVarB}>$
\end{tabular}
\end{center}
}
                        %%%%%%%%%%%%%
               \end{minipage}
        }%\fbox
\end{center}
\caption{Map $\mapMLLToBVT{(\cdot)}$ from formulas and sequents of $\IMLL$ to
structures of $\SBV$}
\label{fig:mapMLLToBVT}
\end{figure}
%%%%%%%%%
\begin{proposition}[\textit{\textbf{Internalizing sequents}}]
\label{proposition:Internalizing sequents}
Let the map $\mapMLLToBVT{(\cdot)}$ from formulas and sequents of $\IMLL$ to
structures of $\SBV$ be given in Figure~\ref{fig:mapMLLToBVT}. Then, we can
extend it to every derivation $\mllPder$ of $\IMLL$ in a way that, if $\mllPder$
has conclusion $\mllA_1,\ldots,\mllA_m\mllJud\mllA$, and \emph{free axioms}
$\mllA_1\mllJud\mllA_1,\ldots,\mllA_m\mllJud\mllA_m$, then:
\[
\vlderivation                             {
\vlde{\mapMLLToBVT{\mllPder}}{}
     {\vlsbr
      <(\vlne{\mapMLLToBVT{\mllA_1}};\vldots;\vlne{\mapMLLToBVT{\mllA_m}})
       ;\mapMLLToBVT{\mllA}>
     }                                    {
\vlhy{\vlsbr
     (<\vlne{\mapMLLToBVT{\mllA_1}};\mapMLLToBVT{\mllA_1}>
      ;\vldots;
      <\vlne{\mapMLLToBVT{\mllA_m}};\mapMLLToBVT{\mllA_m}>)
     }                                    }}
\]
\end{proposition}
\begin{proof}
By induction on the size $\Size{\mllPder}$ of $\mllPder$ which counts the
number of instance rules in it, proceeding by cases on its last rule.
(Details in Appendix~\ref{section:Proof of Internalizing sequents}).
\end{proof}
%%%%%%%%%%%%
Given Proposition~\ref{proposition:Internalizing sequents}, it has been natural
to look for the least extension of $\SBV$ where we could manage the
context-sensitive mechanism of substitution of a term for a variable of
\llcxlinlamcalc\ \llcxwithexs. Such a least extension is the renaming operator
that simply
determines the scope within which we need to search the name that has to be
replaced by (the representation) of a \llcxlinlamterm\ \llcxwithexs.
%%%%%%%%%%%%
\subsection{Properties of the embedding $\mapLcToDi{\cdot}{\cdot}$}
%%%%%%%%
\begin{lemma}[\textbf{\textit{Output names are linear}}]
\label{lemma:Output names are linear}
Every output name of $\mapLcToDi{\llcxM}{o}$ occurs once in it.
\end{lemma}
\begin{proof}
By structural induction on the definition of $\mapLcToDi{\cdot}{\cdot}$,
proceeding by cases on the form of $\llcxM$.
\end{proof}
%%%%%%%%%%%%%%%%%%%%%%%%%%%%%%%%%%%%%%%%%%%%%
\begin{figure}[ht]
\begin{center}
        \fbox{
                \begin{minipage}{.9\textwidth}
		      %%%%%%%%%%%%
		      {\small
\begin{center}
\begin{tabular}{ccc}
$\vlinf{\bvtorerule}{}
       {\vlsbr[\mapLcToDi{\llcxM}{\atmp};(\atmp;\natmo)]}
       { \mapLcToDi{\llcxM}{\atmo}}$
&\quad&
$\vlinf{\bvtsvarule}{}
       {\mapLcToDi{\llcxE{\llcxX}{\llcxX}{\llcxP}}{\atmo}}
       {\mapLcToDi{\llcxP}{\atmo}}$
\\\\
$\vlinf{\bvtsinrule}{}
       {\mapLcToDi{\llcxA{\llcxF{\llcxX}{\llcxM}}{\llcxN}}{\atmo}}
       {\mapLcToDi{\llcxE{\llcxM}{\llcxX}{\llcxN}}{\atmo}}
$
&\quad&
$\vlinf{\bvtslrule}{}
       {\mapLcToDi{\llcxE{\llcxF{\llcxY}{\llcxM}}{\llcxX}{\llcxP}}{\atmo}}
       {\mapLcToDi{\llcxF{\llcxY}{\llcxE{\llcxM}{\llcxX}{\llcxP}}}{\atmo}}
$
 \\\\
$\vlinf{\bvtsalrule}{}
       {\mapLcToDi{\llcxE{\llcxA{\llcxM}{\llcxN}}{\llcxX}{\llcxP}}{\atmo}}
       {\mapLcToDi{\llcxA{\llcxE{\llcxM}{\llcxX}{\llcxP}}{\llcxN}}{\atmo}
        \quad \llcxX\in\llcxFV{\llcxM}}
$
&\quad&
$\vlinf{\bvtsarrule}{}
       {\mapLcToDi{\llcxE{\llcxA{\llcxM}{\llcxN}}{\llcxX}{\llcxP}}{\atmo}}
       {\mapLcToDi{\llcxA{\llcxM}{\llcxE{\llcxN}{\llcxX}{\llcxP}}}{\atmo}
        \quad
        \llcxX\in\llcxFV{\llcxN}}
$
\end{tabular}
\end{center}
}
		      %%%%%%%%%%%%
               \end{minipage}
        }%\fbox
\end{center}
\caption{Derivable rules that simulate $\beta$-reduction \llcxwithexs}
\label{fig:Derivable rules that simulate beta-reduction}
\end{figure}
%%%%%%%%%%%%%%%%%%%%%%%%%%%%%%%%%%%%%%%%%%%%%
\begin{lemma}[\textbf{\textit{Output renaming}}]
\label{lemma:Output renaming}
For every $\llcxM, \atmo$, and  $\atmp$, the rule $\vlrulename{\bvtorerule}$ in
Figure~\ref{fig:Derivable rules that simulate beta-reduction} is derivable in
the down-fragment of $\BVT$.
\end{lemma}
\begin{proof}
By induction on the size of $\llcxM$.
(Details in Appendix~\ref{section:Proof of lemma:Output renaming}).
\end{proof}
%%%%%%%%%%%%%%
\begin{lemma}[\textbf{\textit{Simulating $\llcxBeta$}}]
\label{lemma:Simulating llcxBeta}
For every $\llcxM,\llcxN,\llcxP, \atmo, \atmp$, and  $\atmq$, 
\begin{enumerate}
\item
The rules $\vlrulename{\bvtsinrule}, \vlrulename{\bvtslrule},
\vlrulename{\bvtsalrule}$, and $\vlrulename{\bvtsarrule}$, in
Figure~\ref{fig:Derivable rules that simulate beta-reduction} are derivable in
the down-fragment of $\BVT$.
\item 
The rule $\vlrulename{\bvtsvarule}$ in Figure~\ref{fig:Derivable rules that
simulate beta-reduction} is derivable in the down-fragment of $\BVT$ plus
$\vlrulename{\bvtsequrule}$.
\end{enumerate}
\end{lemma}
\begin{proof}
All the derivations require to apply the above Lemma~\ref{lemma:Output
renaming}. More specifically, $\vlrulename{\bvtsvarule}$ requires
$\vlrulename{\bvtmixprule}$, while $\vlrulename{\bvtsinrule},
\vlrulename{\bvtslrule}, \vlrulename{\bvtsalrule}$, and
$\vlrulename{\bvtsarrule}$ require one instance of $\vlrulename{\bvtrdrule}$.
(Details in Appendix~\ref{section:lemma:Simulating llcxBeta}).
\end{proof}
%%%%%%%%%%%%%%%%%%%
\begin{remark}
Were the clause $\llcxE{\llcxY}{\llcxX}{\llcxP} \llcxBeta \llcxY$ in the
definition of $\llcxBeta$ we could not prove Lemma~\ref{lemma:Simulating
llcxBeta} because we could not prove
$\vlderivation{
\vlde{}{\BVT}{\mapLcToDi{\llcxE{\llcxX}{\llcxX}{\llcxP}}{\atmo}}{ \vlhy
{\mapLcToDi{\llcxP}{\atmo}}}}$. The reason is that, given
$\mapLcToDi{\llcxE{\llcxX}{\llcxX}{\llcxP}}{\atmo}$, it is not evident which
logical tool can erase $\llcxP$ as directly as happens for
$\llcxE{\llcxY}{\llcxX}{\llcxP} \llcxBeta \llcxY$. The only erasure mechanism
existing in $\BVT$ is atom annihilation through the rules
$\vlrulename{\bvtatidrule}$, and $\vlrulename{\bvtatiurule}$.
\end{remark}
%%%%%%%%%%%%%%%%%
\begin{theorem}[\textbf{\textit{Completeness of $\SBVT$}}]
\label{theorem:Completeness of SBVT}
For every $\llcxM$, and  $\atmo$, if $\llcxSOSJud{\llcxM}{\llcxN}$, then
$\bvtInfer{\bvtDder}
          { \mapLcToDi{\llcxN}{\atmo}
            \bvtJudGen{\SBVT}{}
            \mapLcToDi{\llcxM}{\atmo}
 	  }$,
where $\vlrulename{\bvtsequrule}$ is the unique rule of the up-fragment
of $\BVT$ used in $\bvtDder$.
\end{theorem}
\begin{proof}
By induction on $\Size{\llcxSOSJud{\llcxM}{\llcxN}}$, proceeding by cases on the
last rule used, taken among those in Figure~\ref{fig:Rewriting relation
llcxSOSJud}. (Details in Appendix~\ref{section:Proof of Completeness of SBVT}).
\end{proof}
%%%%%%%%%%%%%%%%%
\begin{corollary}[\textbf{\textit{Completeness of $\BVT$}}]
\label{corollary:Completeness of BVT}
For every $\llcxM, \llcxN$, and $\llcxo$, if $\llcxSOSJud{\llcxM}{\llcxN}$, then
$\bvtJudGen{\BVT}{}\vlsbr[\mapLcToDi{\llcxM}{\atmo};
                          \vlne{\mapLcToDi{\llcxN}{\atmo}}]$.
\end{corollary}
\begin{proof}
Theorem~\ref{theorem:Completeness of SBVT}, implies
$\bvtInfer{\bvtDder}{\mapLcToDi{\llcxN}{\atmo} \bvtJudGen{\SBVT}{}
\mapLcToDi{\llcxM}{\atmo}}$.
The rule $\vlrulename{\bvtintdrule}$ is derivable in $\BVT$. So, we can plug it
on top of $\bvtDder$ and apply Theorem~\ref{theorem:Admissibility of the up
fragment} that transforms $\bvtDder$ into some
$\vlstore{\vlsbr[\mapLcToDi{\llcxM}{\atmo}
                   ;\vlne{\mapLcToDi{\llcxN}{\atmo}}]}
 \bvtInfer{\bvtPder}
          {\ \bvtJudGen{\BVT}{}\vlread}
$.
\end{proof}
%%%%%%%%%%%%%%
\begin{theorem}[\textbf{\textit{Soundness of $\SBVT$}}]
\label{theorem:Weak soundness of SBVT}
For every $\llcxM, \llcxN$, and $\llcxo$, let
$\bvtInfer{\bvtDder}{\mapLcToDi{\llcxN}{\atmo} \bvtJudGen{\SBVT}{}
\mapLcToDi{\llcxM}{\atmo} }$ be derived by composing a, possibly empty, sequence
of rules in Figure~\ref{fig:Derivable rules that simulate beta-reduction}.
Then $\llcxSOSJud{\llcxM}{\llcxN}$.
\end{theorem}
\begin{proof}
We reason by induction on $\Size{\bvtDder}$, proceeding by cases on the form of
$\llcxM$ and $\llcxN$.
\par
As a first base case we assume $\llcxM$, and $\llcxN$ coincide. By definition,
every structure is a derivation of $\SBVT$. So, the statement holds.
\par
As a second base case, let $\llcxM$, and $\llcxN$ be different with $\llcxM$ a
redex of $\llcxBeta$ in Figure~\ref{fig:beta-reduction llcxBeta}. So,
$\mapLcToDi{\llcxM}{\llcxo}$ is the conclusion of one of the rules in
Figure~\ref{fig:Derivable rules that simulate beta-reduction}, a part from
$\vlrulename{\bvtorerule}$. We can derive a premise $\mapLcToDi{\llcxN}{\llcxo}$
which, by definition, translates the reduct $\llcxN$. We conclude by
$\vlrulename{\llcxSOSBrule}$ in Figure~\ref{fig:Rewriting relation llcxSOSJud}.
\par
The inductive case is with different $\llcxM$, and $\llcxN$ \ST\ $\llcxM$
contains a redex $\llcxP$. So, $\mapLcToDi{\llcxM}{\atmp} \equiv
\strS\mapLcToDi{\llcxP}{\atmp}$, for some $\strS\vlhole$. As in the previous
case, $\mapLcToDi{\llcxP}{\atmp}$ is the conclusion of some $\bvtrhorule$ among
$\vlrulename{\bvtsinrule}, \vlrulename{\bvtslrule}, \vlrulename{\bvtsalrule},
\vlrulename{\bvtsarrule}$, and $\vlrulename{\bvtsvarule}$. So, it exists
{\small
$\vlderivation                                {
\vlin{\bvtrhorule}{}
     {\mapLcToDi{\llcxM}{\atmp}
      \equiv \strS\mapLcToDi{\llcxP}{\atmp}} {
\vlde{\bvtDder}{}
     {\strS\mapLcToDi{\llcxQ}{\atmp}}       {
\vlhy{\mapLcToDi{\llcxN}{\atmo}}            }}}$
} with $\mapLcToDi{\llcxQ}{\atmp}$ premise of $\bvtrhorule$.
The previous base case in this proof implies $\llcxSOSJud{\llcxP}{\llcxQ}$.
Moreover, $\strS\mapLcToDi{\llcxQ}{\atmp}$ is image,
through $\mapLcToDi{\cdot}{\cdot}$ of
some \llcxlamterm\ $\llcxM'$ since nothing changes in $\strS\vlhole$ when
applying $\bvtrhorule$. Specifically, $\llcxM'$ is $\llcxM$ with
$\llcxQ$ in place of $\llcxP$. So, by induction on
$\bvtInfer{\bvtDder}
          { \mapLcToDi{\llcxN}{\atmo}
             \bvtJudGen{\SBVT}
                       {}
            \mapLcToDi{\llcxM'}{\atmp}
            \equiv \strS\mapLcToDi{\llcxQ}{\atmp}
          } $
we get $\llcxSOSJud{\llcxM'}{\llcxN}$. The conclusion is by an instance
$\vlrulename{\llcxSOStrarule}$ in Figure~\ref{fig:Rewriting relation
llcxSOSJud}.
\end{proof}
%%%%%%%%%%%%%%
\begin{corollary}[\textbf{\textit{Soundness of $\BVT$}}]
\label{corollary:Weak soundness of BVT}
For every $\llcxM, \llcxN$, and $\llcxo$, if
$\vlstore{\vlsbr[\mapLcToDi{\llcxM}{\atmo}
                ;\vlne{\mapLcToDi{\llcxN}{\atmo}}]}
 \bvtInfer{\bvtDder}{\ \bvtJudGen{\BVT}{}\vlread}$, then
$\llcxSOSJud{\llcxM}{\llcxN}$.
\end{corollary}
\begin{proof}
The strategy to build $\bvtDder$ is to start proving
$\bvtInfer{\bvtDder}{\mapLcToDi{\llcxN}{\atmo} \bvtJudGen{\SBVT}{}
\mapLcToDi{\llcxM}{\atmo} }$ in $\SBVT$, for some $\llcxN$, using
Theorem~\ref{theorem:Weak soundness of SBVT}. Then we plug
$\vlrulename{\bvtintdrule}$, which is derivable, on top of $\bvtDder$. Finally,
we apply Corollary~\ref{corollary:Cut elimination}.
\end{proof}
\begin{remark}[\textbf{\textit{Potential of $\BVT$ soundness}}]
\label{remark:Potential of corollary:Weak soundness of BVT}
Corollary~\ref{corollary:Weak soundness of BVT} suggests that proof-search
inside $\BVT$ can be used as an interpreter of \llcxlamcalc\ \llcxwithexs. The
interpreter, however, has a weakness. It works under a specific strategy.
Currently, we do not know if we can reformulate it so that, for example, the
existence of the \emph{shortest proof} of
$\vlsbr[\mapLcToDi{\llcxM}{\atmo};\vlne{\mapLcToDi{\llcxN}{\atmo}}]$ would
always imply that $\llcxM$ evaluates to $\llcxN$. Of course, such a stronger
statement could become relevant in a further extension of $\BVT$ where
\emph{full} \llcxlamcalc\ could be simulated.
\end{remark}
\section{Conclusions and future work}
\label{section:Conclusions and future work}
We define an extension $\SBVT$ of $\SBV$ by introducing an atom renaming
operator which is a self-dual limited version of universal and existential
quantifiers. Renaming and $\vlsbr<\strR;\strT>$ model the evaluation of
\llcxlinlamterms\ \llcxwithexs\ as proof-search in $\SBVT$. So, we have not
applied \DI\ methodology to reformulate an existing logical system we already
know to enjoy a Curry-Howard correspondence with the \llcxlamcalc. Instead, we
have searched to use as much as possible logical operators at the core of \DI,
slightly extended to get a computational behavior we could not obtain
otherwise.
%%%%%%%%%%%%%%%%%
\par
We conclude by listing some of the possible natural developments of the work.
\par
Concerning Remark~\ref{remark:Potential of corollary:Weak soundness of BVT} here
above, extensions of $\SBVT$ whose unconstrained proof-search strategies could
be a sound interpreter of \emph{full} \llcxlamcalc\ (\llcxwithexs) is one
natural work direction. This would really allow to implement one of the
motivations leading to this work, related to the search of new programming
primitives, or evaluation strategies, of (paradigmatic) programming languages.
Starting point to extend $\SBVT$ could be \cite{GuglStra:02:A-Non-co:dq,
StraGugl:09:A-System:vn}.
\par
Also, we can think of extending $\SBVT$ by an operator that models
non-deterministic choice. One reason would be the following generalization of
soundness (Theorem~\ref{theorem:Weak soundness of SBVT},
page \pageref{theorem:Weak soundness of SBVT}).
Let us assume we know that $\llcxM$, applied to $\llcxP$,
reduces to one among $\llcxN_1,\ldots,\llcxN_m$. Proving the statement:
\begin{quote}
If {\small
$\vlsbr[\vlne{\mapLcToDi{\llcxP_1}{\atmo}}
             \vlpl\vldots\vlpl
             \vlne{\mapLcToDi{\llcxP_{i-1}}{\atmo}}
             \vlpl
             \vlne{\mapLcToDi{\llcxP_{i+1}}{\atmo}}
             \vlpl\vldots\vlpl
             \vlne{\mapLcToDi{\llcxP_m}{\atmo}}
            ]
           \bvtJudGen{\BVT}
                     {}
            \vlsbr
          [\mapLcToDi{\llcxM}{\atmo}
          ;[\vlne{\mapLcToDi{\llcxN_1}{\atmo}}
            \vlpl\vldots\vlpl
            \vlne{\mapLcToDi{\llcxN_m}{\atmo}}
           ]
          ]$},
then $\llcxM$ reduces to $\llcxN_i$, for some $\llcxP_1,\ldots,\llcxP_m$.
\end{quote}
\noindent
would represent the evaluation space of any \llcxlinlamterm\ \llcxwithexs\ as a
non-deterministic process searching normal forms. Candidate rules for
non-deterministic choice to extend $\SBVT$ could be\footnote{The conjecture
about the existence of the two rules $\bvtpludrule$, and $\bvtpluurule$, that
model non-deterministic choice, results from discussions with Alessio
Guglielmi}:
{\small
\[
\vlinf{\bvtpludrule}{}
        {{\vlsbr[[\strR\vlpl\strU];\strT]}}
        {{\vlsbr[[\strR;\strT]\vlpl[\strU;\strT]]}}
\qquad\qquad
\vlinf{\bvtpluurule}{}
        {{\vlsbr[(\strR;\strT)\vlpl(\strU;\strT)]}}
        {{\vlsbr([\strR\vlpl\strU];\strT)}}
\]}
\noindent
A further reason to extend $\SBVT$ with non-deterministic choice is to keep
developing the programme started in \cite{Brus:02:A-Purely:wd}, aiming at a
purely logical characterization of \emph{full} $\CCS$. We recall that in
\cite{Brus:02:A-Purely:wd} only sequential and parallel composition of processes
have been casted in logical terms.
\par
Finally, the exploration of relations between linear \llcxlamcalc\ \llcxwithexs,
as we embed it in $\SBVT$ using a calculus-of-process style, and the evolution
of quantum systems, as proofs of $\BVT$ \cite{BlutGuglIvanPana:10:A-Logica:uq},
makes sense. Indeed, modeling a $\lambda$-variable $\llcxX$ as a forwarder
$\vlsbr<\llcxX;\natmo>$ is, essentially, looking at $\llcxX$ as a sub-case of
$\vlsbr<(\llcxX_1;\vldots;\llcxX_k);[\natmo_1;\vldots;\natmo_l]>$, which recalls
the origins of our embedding and which can represent edges in DAGs
that model quantum systems evolution \cite{BlutGuglIvanPana:10:A-Logica:uq}.
%%%%%%%%%%%%%
% \input{BV2Min}
% \input{SBV2Min-to-LLCXS}
%%%%%%%%%%%%%%%%%%%%%%%%%%%%%%%%%%%%%%%%%%%%%%%%%%%%%%%%%%%%%%%%%%%%%%%%%%
% \bibliographystyle{alphaurl}
\bibliographystyle{plain}
\bibliography{10-LinLambdaCalc-in-di}
%%%%%%%%%%%%%%%%%%%%%%%%%%%%%%%%%%%%%%%%%%%%%%%%%%%%%%%%%%%%%%%%%%%%%%%%%%
\appendix
\section{Proof of \textit{\textbf{Context extrusion}}
(Proposition~\ref{proposition:Context extrusion},
page~\pageref{proposition:Context extrusion})}
\label{section:Proof of proposition:Context extrusion}
By induction on
$\Size{\strS\vlhole}$, proceeding by cases on the form of
$\strS\vlhole$. In the base case with $\strS\vlhole\equiv\vlhole$, the statement
holds simply because
$\strS\vlsbr[\strR;\strT]
 \equiv\vlsbr[\strS\vlscn{\strR};\strT]$,
and $\strS\vlsbr[\strR;\strT]$, being it a structure,
is, by definition, a derivation.
%%%% first case
\par
As a \emph{first case}, let $\strS\vlhole\equiv {\vlsbr<\strS'\vlhole;\strU>}$.
Then:
\[
\vlderivation                                            {
\vliq{\bvtseqdrule}{}
     {\vlsbr[\strS\vlscn{\strR};\strT]
      \equiv\vlsbr[<\strS'\vlscn{\strR};\strU>;\strT]} {
\vlde{\bvtDder}{}
     {\vlsbr<[\strS'\vlscn{\strR};\strT];\strU>}      {
\vlhy{\vlsbr<\strS'\vlscn{[\strR;\strT]};\strU>
      \equiv\strS\vlsbr[\strR;\strT]                  }}}}
\]
where $\bvtDder$ exists by inductive hypothesis which holds thanks to
$\Size{\strS'\vlhole}<\Size{\strS\vlhole}$.
If, instead $\strS\vlhole\equiv {\vlsbr(\strS'\vlhole;\strU)}$,
we can proceed as here above, using $\vlrulename{\bvtswirule}$ in place of
$\vlrulename{\bvtseqdrule}$.
%%%% second case
\par
As a \emph{second case}, let $\strS\vlhole\equiv\vlfo{\atma}{\strS'\vlhole}$.
Then:
\[
\vlderivation                                                 {
\vliq{}{}
     {\vlsbr[\strS\vlscn{\strR};\strT]
      \equiv\vlsbr[\vlfo{\atma}{\strS'\vlscn{\strR}};\strT]} {
\vliq{\bvtrdrule}{}
     {\vlsbr[\vlfo{\atma}{\strS'\vlscn{\strR}}
            ;\vlfo{\atma}{\strT\subst{\atma}{\atma}}]}   {
\vlde{\bvtDder}{}
     {\vlfo{\atma}{\vlsbr[\strS'\vlscn{\strR};\strT]}}  {
\vlhy{\vlfo{\atma}{\strS'\vlsbr[\strR;\strT]         }
      \equiv\strS\vlsbr[\strR;\strT]                   }}}}}
\]
where $\bvtDder$ exists by inductive hypothesis which holds thanks to
$\Size{\strS'\vlhole}<\Size{\strS\vlhole}$.

\section{Proof of \textit{\textbf{Shallow
Splitting}} (Proposition~\ref{proposition:Shallow Splitting},
page~\pageref{proposition:Shallow Splitting})}
\label{section:Proof of proposition:Shallow Splitting}
Point~\ref{enum:Shallow-Splitting-atom} holds by starting to observe that $\strP
\not\approx\vlone$. Otherwise, we would contradict the assumption. Then, we
proceed by induction on $\Size{\bvtDder}$, reasoning by cases on the last rule
$\bvtrhorule$ of $\bvtDder$. If $\bvtrhorule$ is $\vlrulename{\bvtatidrule}$
then $\strP$ is
$\natma$. Otherwise, $\bvtrhorule$ rewrites $\strP$ to some $\strP'$, getting
$\vlstore{\vlsbr[\atma;\strP']} \bvtInfer{\bvtDder'}{\ \bvtJ\vlread}$, which, by
inductive hypothesis, implies $\bvtInfer{\bvtDder'}{\natma \bvtJ \strP'}$. The
application of $\bvtrhorule$ gives the thesis.
\par
From \cite{Gugl:06:A-System:kl} we know that the statements
\ref{enum:Shallow-Splitting-seq} and \ref{enum:Shallow-Splitting-copar} hold in
$\BV$ by induction on the lexicographic order of the pair $(\Size{\strV},
\Size{\bvtDder})$, where $\strV$ is one between ${\vlsbr[<\strR;\strT>;\strP]}$
or ${\vlsbr[(\strR;\strT);\strP]}$, proceeding by cases on the last rule
$\bvtrhorule$ of $\bvtDder$.
\par
We start extending the proof of points~\ref{enum:Shallow-Splitting-seq},
and~\ref{enum:Shallow-Splitting-copar} to the cases where $\bvtrhorule$ is
$\vlrulename{\bvtrdrule}$, hence proving that
points~\ref{enum:Shallow-Splitting-seq},
and~\ref{enum:Shallow-Splitting-copar} hold inside $\BVT$. We focus on
point~\ref{enum:Shallow-Splitting-seq}, being~\ref{enum:Shallow-Splitting-copar}
analogous.
\par
Let the last rule of $\bvtDder$ be $\vlrulename{\bvtrdrule}$. If its redex falls
inside
$\strR, \strT$ or $\strP$ it is enough to proceed by induction on
$\Size{\bvtDder}$. Otherwise, the redex of $\vlrulename{\bvtrdrule}$ can be the
whole
$\vlsbr[<\strR;\strT>;\strP]$, thanks to
$\vlsbr[<\strR;\strT>;\strP]\approx
\vlstore{\vlfo{\atma}{\vlsbr<\strR;\strT>};\vlfo{\atma}{\strP}} \vlsbr[\vlread]
$, if
$\vlstore{\vlsbr<\strR;\strT>}
 \atma\not\in\strFN{\vlread}\cup\strFN{\strP}$.
So we have
\[
\vlderivation                                                     {
\vlin{\bvtrdrule}{}
     {\vlsbr[\vlfo{\atma}{<\strR;\strT>};\vlfo{ \atma }{\strP}]}{
\vlpr{\bvtDder'}{}
     {\vlsbr[\vlfo{\atma}{<\strR;\strT>;\strP}]                }}}
\]
The derivability of structures (Proposition~\ref{proposition:Derivability of
structures in BVT}) applied on $\bvtDder'$ implies that, for every $\atmb$,
\[
\vlderivation                                           {
\vliq{}{}
     {\vlsbr[<\strR;\strT>;\strP]\subst{\atmb}{\atma} }{
\vlpr{\bvtDder''}{}
     {\vlsbr[<\strR;\strT>;\strP]                     }}}
\]
The inductive hypothesis holds thanks to $\Size{\bvtDder''}<\Size{\bvtDder}$.
So, there are $\strP', \strP''$ \ST\
$\vlsbr<\strP';\strP''> \bvtJ \strP$,
$\bvtJ {\vlsbr[\strR;\strP']}$, and
$\bvtJ {\vlsbr[\strT;\strP'']}$, which prove the statement.
\par
Now, we prove point~\ref{enum:Shallow-Splitting-fo} by detailing the three
relevant cases.
%%%% first case of shallow splitting, point enum:Shallow-Splitting-fo
\par\noindent
As a \emph{first case}, let $P$ be $\vlsbr[<\strT';\strT''>;\strV';\strV'']$.
So $\bvtDder$ can
be:
\[
\vlderivation                                                                  {
\vliq{          }{}
     {\vlsbr[\vlfo{\atma}{\strR};<\strT';\strT''>;\strV';\strV'']}            {
\vlin{\bvtseqdrule}{}
     {\vlsbr[<\vlone;[\vlfo{\atma}{\strR};\strV']>;<\strT';\strT''>;\strV'']}{
\vliq{          }{}
     {\vlsbr[<[\vlone;\strT'];[\vlfo{\atma}{\strR};\strV';\strT'']>;\strV'']}{
\vlpr{\bvtDder'}{}
     {\vlsbr[<\strT';[\vlfo{\atma}{\strR};\strV';\strT'']>;\strV'']}}}}}
\]
The relations
$\vlstore{\vlsbr[<\strT';[\vlfo{\atma}{\strR};\strV';\strT'']>;\strV'']}
 \Size{\vlread}
  =
 \vlstore{\vlsbr[\vlfo{\atma}{\strR};<\strT';\strT''>;\strV';\strV'']}
 \Size{\vlread}$, and
$\Size{\bvtDder'} < \Size{\bvtDder}$ imply the inductive hypothesis holds for
point~\ref{enum:Shallow-Splitting-seq} on $\bvtDder'$. So,
there are
$\strP', \strP''$ \ST\
$\vlstore{\vlsbr<\strP';\strP''>}
 \bvtInfer{\bvtDder''}{\vlread \bvtJ \strV''}$,
$\vlstore{\vlsbr[\strT';\strP']}
 \bvtInfer{\bvtDder'''}{\ \bvtJ \vlread}$, and
$\vlstore{\vlsbr[\vlfo{\atma}{\strR};\strV';\strT'';\strP'']}
 \bvtInfer{\bvtPder}{\ \bvtJ {\vlread}}$.
The relation $\Size{\strP''} < \Size{\strV''}$ implies that
$\vlstore{\vlsbr[\vlfo{\atma}{\strR};\strV';\strT'';\strP'']}
 \Size{{\vlread}}
  <
 \vlstore{\vlsbr[\vlfo{\atma}{\strR};<\strT';\strT''>;\strV';\strV'']}
 \Size{{\vlread}}$. So, the inductive
hypothesis holds for point~\ref{enum:Shallow-Splitting-fo} on $\bvtPder$. Hence,
there is
$\strR'$ \ST\
$\vlstore{\vlsbr[\strV';\strT'';\strP'']}
 \bvtInfer{\bvtPder'}{\vlex{\atma}{\strR'} \bvtJ{\vlread}}$, and
$\vlstore{\vlsbr[\strR;\strR']}
 \bvtInfer{\bvtPder''}{\ \bvtJ {\vlread}}$, where $\bvtPder''$ is the
``second half'' of our thesis. Instead, the ``first half'' is:
\[
\vlderivation                                                                {
\vliq{          }{}
     {\vlsbr[<\strT';\strT''>;\strV';\strV'']}                               {
\vlin{\bvtseqdrule}{}
     {\vlsbr[\strV'';[<\vlone;\strV'>;<\strT';\strT''>]]}                    {
\vliq{          }{}
     {\vlsbr[\strV'';<[\vlone;\strT'];[\strV';\strT'']>]}                    {
\vlde{\bvtDder''}{}
     {\vlsbr[\strV'';<\strT';[\strV';\strT'']>]}                             {
\vlin{\bvtseqdrule}{}
     {\vlsbr[<\strP';\strP''>;<\strT';[\strV';\strT'']>]}                    {
\vlde{\vlsbr<\bvtDder''';\,\bvtPder'>}{}
     {\vlsbr<[\strT';\strP'];[\strV';\strT'';\strP'']>}                      {
\vliq{          }{}
     {\vlsbr<\vlone;\vlfo{\atma}{\strR'}>}                                   {
\vlhy{\vlfo{\atma}{\strR'}}                                           }}}}}}}}
\]
%%%% second case of shallow splitting, point enum:Shallow-Splitting-fo
\par\noindent
As a \emph{second case}, let $P$ be $\vlsbr[(\strT';\strT'');\strV';\strV'']$.
So $\bvtDder$ can
be:
\[
\vlderivation                                                                {
\vliq{          }{}
     {\vlsbr[\vlfo{\atma}{\strR};(\strT';\strT'');\strV';\strV'']}           {
\vlin{\bvtswirule}{}
     {\vlsbr[[\vlfo{\atma}{\strR};\strV'];(\strT';\strT'');\strV'']}         {
\vlpr{\bvtDder' }{}
     {\vlsbr[(\strT';[\vlfo{\atma}{\strR};\strV';\strT'']);\strV'']}       }}}
\]
The relations
$\vlstore{\vlsbr[(\strT';[\vlfo{\atma}{\strR};\strV';\strT'']);\strV'']}
 \Size{{\vlread}}
  =
 \vlstore{\vlsbr[\vlfo{\atma}{\strR};(\strT';\strT'');\strV';\strV'']}
 \Size{{\vlread}}$, and
$\Size{\bvtDder'} < \Size{\bvtDder}$ imply the inductive hypothesis holds for
point~\ref{enum:Shallow-Splitting-copar} on $\bvtDder'$. So, there are
$\strP', \strP''$ \ST\
$\vlstore{\vlsbr[\strP';\strP'']}
 \bvtInfer{\bvtDder''}{{\vlread} \bvtJ \strV''}$,
$\vlstore{\vlsbr[\strT';\strP']}
 \bvtInfer{\bvtDder'''}{\ \bvtJ {\vlread}}$, and
$\vlstore{\vlsbr[\vlfo{\atma}{\strR};\strV';\strT'';\strP'']}
\bvtInfer{\bvtPder}{\ \bvtJ{\vlread}}$.
The relation $\Size{\strP''}< \Size{\strV''}$ implies
$\vlstore{\vlsbr[\vlfo{\atma}{\strR};\strV';\strT'';\strP'']}
 \Size{{\vlread}}
  <
 \vlstore{\vlsbr[\vlfo{\atma}{\strR};(\strT';\strT'');\strV';\strV'']}
 \Size{{\vlread}}$. So, the inductive hypothesis holds for
point~\ref{enum:Shallow-Splitting-fo} on $\bvtPder$. Hence,
there is
$\strR'$ \ST\
$\vlstore{\vlsbr[\strV';\strT'';\strP'']}
 \bvtInfer{\bvtPder'}{\vlex{\atma}{\strR'} \bvtJ {\vlread}}$, and
$\vlstore{\vlsbr[\strR;\strR']}
 \bvtInfer{\bvtPder''}{\ \bvtJ {\vlread}}$, where $\bvtPder''$ is the
``second half'' of our thesis. Instead, the ``first half'' is:
\[
\vlderivation                                                 {
\vlde{\bvtDder''}{}
     {\vlsbr[(\strT';\strT'');\strV';\strV'']}                {
\vliq{          }{}
     {\vlsbr[(\strT';\strT'');\strV';[\strP';\strP'']]}       {
\vlin{\bvtswirule}{}
     {\vlsbr[[(\strT';\strT'');\strP'];\strV';\strP'']}       {
\vlde{\bvtDder'''}{}
     {\vlsbr[([\strT';\strP'];\strT'');\strV';\strP'']}       {
\vliq{          }{}
     {\vlsbr[(\vlone;\strT'');\strV';\strP'']}                {
\vlde{\bvtPder'}{}
     {\vlsbr[\strV';\strT'';\strP'']}                         {
\vlhy{\vlfo{\atma}{\strR'}}                                   }}}}}}}
\]
%%%% third case of shallow splitting, point enum:Shallow-Splitting-fo
\par\noindent
As a \emph{third case}, let $P$ be $\vlsbr[\vlfo{\atma}{\strT'};\strT'']$. So
$\bvtDder$ can be:
\[
\vlderivation                                                       {
\vlin{\bvtrdrule}{}
     {\vlsbr[\vlfo{\atma}{\strR};\vlfo{\atma}{\strT'};\strT'']}     {
\vlpr{\bvtDder' }{}
     {\vlsbr[\vlfo{\atma}{[\strR;\strT']};\strT'']}                 }}
\]
The relation
$\vlstore{\vlsbr[\vlfo{\atma}{[\strR;\strT']};\strT'']}
 \Size{{\vlread}}
 \leq
 \vlstore{\vlsbr[\vlfo{\atma}{\strR};\vlfo{\atma}{\strT'};\strT'']}
 \Size{{\vlread}}$, and $\Size{\bvtDder'}<\Size{\bvtDder}$
implies the inductive hypothesis holds for point~\ref{enum:Shallow-Splitting-fo}
on $\bvtDder'$. So, there is $\strR'$ \ST\
$\bvtInfer{\bvtPder}{\vlex{\atma}{\strR'} \bvtJ \strT''}$, and
$\vlstore{\vlsbr[\strR;\strT';\strR']}
 \bvtInfer{\bvtPder'}{\ \bvtJ {\vlread}}$, where
$\bvtPder'$ is the
``second half'' of our thesis. Instead, the ``first half'' is:
\[
\vlderivation                                                 {
\vlde{\bvtPder}{}
     {\vlsbr[\vlfo{\atma}{\strT'};\strT'']}                   {
\vlin{\bvtrdrule}{}
     {\vlsbr[\vlfo{\atma}{\strT'};\vlfo{\atma}{\strR'}]}      {
\vlhy{\vlfo{\atma}{\vlsbr[\strT';\strR']}}                    }}}
\]

\section{Proof of \textit{Context
Reduction} (Proposition~\ref{proposition:Context Reduction},
page~\pageref{proposition:Context Reduction})}
\label{section:Proof of proposition:Context Reduction}
The proof is by induction on $\Size{\strS\vlhole}$, proceeding by cases on the
form of $\strS\vlhole$.
%%%% first case of context reduction
\par\noindent
As a \emph{first case}, let $\strS\vlhole\equiv \vlsbr<\strS'\vlhole;\strP>$.
So, the assumption is
$\vlstore{\vlsbr<\strS'\vlscn\strR;\strP>}
 \bvtInfer{\bvtDder}{\ \bvtJ{\vlread}}$.
The derivability
of structures implies $\bvtInfer{\bvtDder'}{\ \bvtJ \strS'\vlscn{\strR}}$, and
$\bvtInfer{\bvtDder''}{\ \bvtJ \strP}$.
The relation
$\Size{\strS'\vlscn{\strR}} <
 \vlstore{\vlsbr<\strS'\vlscn{\strR};\strP>}
 \Size{\vlread}$
implies the inductive hypothesis holds on $\bvtDder'$.
So, there are $\strU, \atma$ \ST\
$\vlstore{\vlfo{\atma}{\vlsbr[\vlhole;\strU]}}
 \bvtInfer{\bvtPder}{\vlread \bvtJ \strS'\vlhole}$,
and $\bvtJ {\vlsbr[\strR;\strU]}$ which is the ``second half'' of the thesis.
Instead, the ``first half'' is:
\[
\vlderivation                                   {
\vlde{\bvtDder''}{}
     {\vlsbr<\strS'\vlhole;\strP>}              {
\vliq{}{}
     {\vlsbr<\strS'\vlhole;\vlone>}             {
\vlde{\bvtPder}{}
     {\strS'\vlhole}                            {
\vlhy{\vlfo{\atma}{\vlsbr[\vlhole;\strU]}}      }}}}
\]
%%%% second case of context reduction
\par\noindent
As a \emph{second case}, let $\strS\vlhole\equiv\vlsbr(\strS'\vlhole;\strP)$.
So, the assumption is
$\vlstore{\vlsbr(\strS'\vlscn\strR;\strP)}
 \bvtInfer{\bvtDder}{\ \bvtJ{\vlread}}$. The derivability
of structures implies $\bvtInfer{\bvtDder'}{\ \bvtJ \strS'\vlscn{\strR}}$, and
$\bvtInfer{\bvtDder''}{\ \bvtJ \strP}$.
The relation
$\Size{\strS'\vlscn{\strR}} <
 \vlstore{\vlsbr(\strS'\vlscn{\strR};\strP)}
 \Size{\vlread}$
implies the inductive hypothesis holds on $\bvtDder'$.
So, there are $\strU, \atma$ \ST\
$\vlstore{\vlfo{\atma}{\vlsbr[\vlhole;\strU]}}
 \bvtInfer{\bvtPder}{\vlread \bvtJ \strS'\vlhole}$,
and $\bvtJ {\vlsbr[\strR;\strU]}$ which is the ``second half'' of the thesis.
Instead, the ``first half'' is:
\[
\vlderivation                                   {
\vlde{\bvtDder''}{}
     {\vlsbr(\strS'\vlhole;\strP)}              {
\vliq{}{}
     {\vlsbr(\strS'\vlhole;\vlone)}             {
\vlde{\bvtPder}{}
     {\strS'\vlhole}                            {
\vlhy{\vlfo{\atma}{\vlsbr[\vlhole;\strU]}}      }}}}
\]
%%%% third case of context reduction
\par\noindent
As a \emph{third case}, let $\strS\vlhole\equiv\vlfo{\atma}{\strS'\vlhole}$
with $\atma\in\strFN{\strS'\vlscn{\strR}}$. Otherwise, we have to consider the
case suitable to treat $\strS\vlhole\equiv\strS'\vlhole\approx{\vlfo{\atma}{
\strS'\vlhole}}$.
So, the assumption is $\bvtInfer{\bvtDder}{\ \bvtJ
\vlfo{\atma}{\strS'\vlscn{\strR}}}$. The derivability
of structures (Proposition~\ref{proposition:Derivability of structures in
BVT}) implies $\bvtInfer{\bvtDder'}{\ \bvtJ \strS'\vlscn{\strR}}$.
The relation
$\Size{\strS'\vlscn{\strR}} < \Size{\vlfo{\atma}{\strS'\vlscn{\strR}}}$
implies the inductive hypothesis holds $\bvtDder'$.
So, there are $\strU, \atma$ \ST\
$\vlstore{\vlfo{\atma}{\vlsbr[\vlhole;\strU]}}
 \bvtInfer{\bvtPder}{\vlread \bvtJ \strS'\vlhole}$,
and $\bvtJ {\vlsbr[\strR;\strU]}$ which is the ``second half'' of the thesis.
Instead, the ``first half'' is:
\[
\vlderivation                                                 {
\vlde{\vlfo{\atma}{\bvtPder}}{}
     {\vlfo{\atma}{\strS'\vlhole}}                            {
\vliq{\bvtrdrule}{}
     {\vlfo{\atma}{\vlfo{\atma}{\vlsbr[\vlhole;\strU]}}}      {
\vlhy{\vlfo{\atma}{\vlsbr[\vlhole;\strU]}}                    }}}
\]
%%%% fourth case of context reduction
\par\noindent
As a \emph{fourth case}, let
$\strS\vlhole\equiv\vlsbr[<\strS''\vlhole;\strP'>;\strP]$.
So, the assumption is
$\vlstore{\vlsbr[<\strS''\vlscn\strR;\strP'>;\strP]}
 \bvtInfer{\bvtDder}{\ \bvtJ{\vlread}}$.
Shallow splitting implies the existence of
$\strP_1, \strP_2$ \ST\
$\vlstore{\vlsbr<\strP_1;\strP_2>}
 \bvtInfer{\bvtPder}{{\vlread} \bvtJ \strP}$,
$\vlstore{\vlsbr[\strS''\vlscn\strR;\strP_1]}
 \bvtInfer{\bvtPder_0}{\ \bvtJ {\vlread}}$, and
$\vlstore{\vlsbr[\strP';\strP_2]}
 \bvtInfer{\bvtPder_1}{\ \bvtJ {\vlread}}$.
The relation
$\vlstore{\vlsbr[\strS''\vlscn\strR;\strP_1]}
 \Size{{\vlread}} <
 \vlstore{\vlsbr[<\strS''\vlscn\strR;\strP'>;\strP]}
 \Size{{\vlread}}$, which holds also
thanks to
$\Size{\strP_1}<\Size{\strP}$,
implies the inductive hypothesis holds on $\bvtPder_0$.
So, there are $\strU, \atma$ \ST\
$\vlstore{\vlfo{\atma}{\vlsbr[\vlhole;\strU]} \bvtJ
  {\vlsbr[\strS''\vlhole;\strP_1]}
 } \bvtInfer{\bvtPder'}{\vlread}$,
and $\bvtJ {\vlsbr[\strR;\strU]}$ which is the ``second half'' of the thesis.
Instead, the ``first half'' is:
\[
\vlderivation                                            {
\vlde{\bvtPder}{}
     {\vlsbr[<\strS''\vlhole;\strP'>;\strP]}             {
\vlin{\bvtseqdrule}{}
     {\vlsbr[<\strS''\vlhole;\strP'>;<\strP_1;\strP_2>]} {
\vlde{\bvtPder_1}{}
     {\vlsbr<[\strS''\vlhole;\strP_1];[\strP';\strP_2]>} {
\vliq{}{}
     {\vlsbr<[\strS''\vlhole;\strP_1];\vlone>}           {
\vlde{\bvtPder'}{}
     {\vlsbr[\strS''\vlhole;\strP_1]}                    {
\vlhy{\vlfo{\atma}{\vlsbr[\vlhole;\strU]}}               }}}}}}
\]
%%%% fifth case of third of context reduction
\par\noindent
As a \emph{fifth case}, let
$\strS\vlhole\equiv\vlsbr[(\strS''\vlscn\strR;\strP');\strP]$.
So, the assumption is
$\vlstore{\vlsbr[(\strS''\vlscn\strR;\strP');\strP]}
\bvtInfer{\bvtDder}{\ \bvtJ{\vlread}}$.
Shallow splitting implies the existence of
$\strP_1, \strP_2$ \ST\
$\vlstore{\vlsbr[\strP_1;\strP_2]}
 \bvtInfer{\bvtPder}{{\vlread} \bvtJ \strP}$,
$\vlstore{\vlsbr[\strS''\vlscn\strR;\strP_1]}
 \bvtInfer{\bvtPder_0}{\ \bvtJ {\vlread}}$,
and
$\vlstore{\vlsbr[\strP';\strP_2]}
 \bvtInfer{\bvtPder_1}{\ \bvtJ {\vlread}}$.
The relation
$\vlstore{\vlsbr[\strS''\vlscn\strR;\strP_1]}
 \Size{{\vlread}} <
 \vlstore{\vlsbr[(\strS''\vlscn\strR;\strP');\strP]}
 \Size{{\vlread}}$, which holds also thanks to
$\Size{\strP_1}<\Size{\strP}$,
implies the inductive hypothesis holds on $\bvtPder_0$.
So, there are $\strU, \atma$ \ST\
$\vlstore{\vlfo{\atma}{\vlsbr[\vlhole;\strU]}
           \bvtJ {\vlsbr[\strS''\vlhole;\strP_1]}}
 \bvtInfer{\bvtPder'}{\vlread}$,
and $\bvtJ {\vlsbr[\strR;\strU]}$ which is the ``second half'' of the thesis.
Instead, the ``first half'' is:
\[
\vlderivation                                            {
\vlde{\bvtPder}{}
     {\vlsbr[(\strS''\vlhole;\strP');\strP]}             {
\vliq{}{}
     {\vlsbr[(\strS''\vlhole;\strP');[\strP_1;\strP_2]]} {
\vlin{\bvtswirule}{}
     {\vlsbr[[(\strS''\vlhole;\strP');\strP_1];\strP_2]} {
\vlin{\bvtswirule}{}
     {\vlsbr[([\strS''\vlhole;\strP_1];\strP');\strP_2]} {
\vlde{\bvtPder_1}{}
     {\vlsbr([\strP';\strP_2];[\strS''\vlhole;\strP_1])} {
\vliq{}{}
     {\vlsbr(\vlone;[\strS''\vlhole;\strP_1])}           {
\vlde{\bvtPder'}{}
     {\vlsbr[\strS''\vlhole;\strP_1]}                    {
\vlhy{\vlfo{\atma}{\vlsbr[\vlhole;\strU]}}               }}}}}}}}
\]
%%%% sixth case of third of context reduction
\par\noindent
As a \emph{sixth case}, let
$\vlstore{\vlsbr[\vlfo{\atma}{\strS''\vlhole};\strP]}
 \strS\vlhole\equiv\vlread$ with
$\atma\in\strFN{\strS''\vlscn{\strR}}$. Otherwise, we have to consider the
case suitable to treat
$\vlstore{\vlsbr[\strS''\vlhole;\strP]}
 \strS\vlhole\equiv{\vlread}\approx
 \vlstore{\vlsbr[\vlfo{\atma}{\strS''\vlhole};\strP]}
 {\vlread}$.
So, the assumption is
$\vlstore{\vlsbr[\vlfo{\atma}{\strS''\vlscn\strR};\strP]}
 \bvtInfer{\bvtDder}{\ \bvtJ
 {\vlread}}$.
Shallow splitting implies the existence of
$\strP'$ \ST\
$\vlex{\atma}{\strP'} \bvtJ \strP$, and
$\vlstore{\vlsbr[\strS''\vlscn\strR;\strP']}
 \bvtInfer{\bvtPder}{\ \bvtJ {\vlread}}$.
The relation
$\vlstore{\vlsbr[\strS''\vlscn\strR;\strP']}
 \Size{{\vlread}} <
 \vlstore{\vlsbr[\vlfo{\atma}{\strS''\vlscn\strR};\strP]}
 \Size{{\vlread}}$
implies the inductive hypothesis holds on $\bvtPder$.
So, there are $\strU, \atma$ \ST\
$\vlstore{\vlfo{\atma}{\vlsbr[\vlhole;\strU]}
          \bvtJ {\vlsbr[\strS''\vlhole;\strP']}}
 \bvtInfer{\bvtPder'}{\vlread}$,
and $\bvtJ {\vlsbr[\strR;\strU]}$ which is the ``second half'' of the thesis.
For getting to the ``first half'' we start observing that
$\bvtPder'$ implies
$\vlstore{\vlex{\atma}{\vlfo{\atma}{\vlsbr[\vlhole;\strU]}}
           \bvtJ
           \vlfo{\atma}{\vlsbr[\strS''\vlhole;\strP']}}
 \bvtInfer{\bvtPder''}{\vlread}$ and
that
$\vlstore{\vlfo{\atma}{\vlsbr[\vlhole;\strU]}}
 \vlex{\atma}{\vlread}
 \approx
 \vlstore{\vlsbr[\vlhole;\strU]}
 \vlfo{\atma}{\vlread}$. So:
\[
\vlderivation                                                    {
\vlde{\bvtPder}{}
     {\vlsbr[\vlfo{\atma}{\strS''\vlhole};\strP]}                {
\vlin{\bvtrdrule}{}
     {\vlsbr[\vlfo{\atma}{\strS''\vlhole};\vlfo{\atma}{\strP'}]} {
\vlde{\bvtPder''}{}
     {\vlfo{\atma}{\vlsbr[\strS''\vlhole;\strP']}}               {
\vlhy{\vlfo{\atma}{\vlsbr[\vlhole;\strU]}}                    }}}}
\]
\section{Proof of \textit{Splitting} (Theorem~\ref{theorem:Splitting},
page~\pageref{theorem:Splitting})}
\label{section:Proof of theorem:Splitting}
We obtain the proof of the three statements by composing Context Reduction
(Proposition~\ref{proposition:Context Reduction}), and Shallow Splitting
(Proposition~\ref{proposition:Shallow Splitting}) in this order.
\par
We give the details of points~\ref{enum:Splitting-seq},
and~\ref{enum:Splitting-fo-ex}, as point~\ref{enum:Splitting-copar} is analogous
to~\ref{enum:Splitting-seq}.
%%% first case
\par
As a \textit{first case}, let us focus on point~\ref{enum:Splitting-seq}.
Context Reduction (Proposition~\ref{proposition:Context Reduction}) applies to
$\bvtDder$. So, there are $\strU, \atma$ \ST\
$\vlstore{\vlsbr[\vlhole;\strU]}
 \bvtInfer{\bvtPder_0}{\vlfo{\atma}{\vlread} \bvtJ \strS\vlhole}$,
and $\vlstore{\vlsbr[<\strR;\strT>;\strU]}
     \bvtInfer{\bvtPder_1}{\ \bvtJ {\vlread}}$. Shallow
Splitting (Proposition~\ref{proposition:Shallow Splitting}) applies to
$\bvtPder_1$.
So,
$\vlstore{\vlsbr<\strK_1;\strK_2>}
 \bvtInfer{\bvtDder_0}{{\vlread} \bvtJ \strU}$,
$\vlstore{\vlsbr[\strR;\strK_1]}
 \bvtInfer{\bvtDder_1}{\ \bvtJ {\vlread}}$, and
$\vlstore{\vlsbr[\strT;\strK_2]}
 \bvtInfer{\bvtDder_2}{\ \bvtJ {\vlread}}$, for some
$\strK_1, \strK_2$. Both $\bvtDder_1$, and $\bvtDder_2$ are the ``second half''
of the proof. The ``first half'' is:
\[
\vlderivation                                           {
\vlde{\bvtPder_0}{}
     {\strS\vlhole                                   } {
\vlde{\bvtDder_0}{}
     {\vlfo{\atma}{\vlsbr[\vlhole;\strU]}            }{
\vlhy{\vlfo{\atma}{\vlsbr[\vlhole;<\strK_1;\strK_2>]}}}}}
\]
%%% second case
\par
As a \textit{second case}, let us focus on point~\ref{enum:Splitting-fo-ex}.
Context Reduction (Proposition~\ref{proposition:Context Reduction}) applies to
$\bvtDder$. So, there are $\strU, \atma$ \ST\
$\vlstore{\vlsbr[\vlhole;\strU]}
 \bvtInfer{\bvtPder_0}{\vlfo{\atma}{\vlread} \bvtJ \strS\vlhole}$,
and $\vlstore{\vlsbr[\vlfo{\atma}{\strR};\strU]}
     \bvtInfer{\bvtPder_1}{\ \bvtJ {\vlread}}$.
We notice that the existence of $\bvtPder_0$ means that, for every $\strV$,
$\vlstore{\vlsbr[\strV;\strU]}
 \bvtInfer{\bvtPder_0}{\vlfo{\atma}{\vlread} \bvtJ \strS\vlscn{\strV}}$.
Shallow Splitting (Proposition~\ref{proposition:Shallow
Splitting}) applies to $\bvtPder_1$.
So,
$\bvtInfer{\bvtDder_0}{\vlfo{\atma}{\strK} \bvtJ \strU}$,
$\vlstore{\vlsbr[\strR;\strK]}
 \bvtInfer{\bvtDder_1}{\ \bvtJ {\vlread}}$, for some
$\strK$. So, $\bvtDder_1$ is the ``second half''
of the proof. For every $\strV$, the ``first half'' is:
\[
\vlderivation                                           {
\vlde{\bvtPder_0}{}
     {\strS\vlscn{\strV}                             } {
\vlde{\bvtDder_0}{}
     {\vlfo{\atma}{\vlsbr[\strV;\strU]}              }{
\vliq{}{}
     {\vlfo{\atma}{\vlsbr[\strV
                         ;\vlfo{\atma}{\strK}
                         ]}                          }{
\vlin{\bvtrdrule}{}
     {\vlfo{\atma}
           {\vlsbr[\vlfo{\atma}{\strV\subst{\atma}{\atma}}
                  ;\vlfo{\atma}{\strK}
                  ]}                                 }{
\vliq{}{}
     {\vlfo{\atma}
      {\vlfo{\atma}{\vlsbr[\strV\subst{\atma}{\atma}
                          ;\strK
                          ]}}                        }{
\vlhy{\vlfo{\atma}{\vlsbr[\strV;\strK]}              }}}}}}}
\]
\section{Proof of \textit{Admissibility of the up fragment}
(Theorem~\ref{theorem:Admissibility of the up fragment},
page~\pageref{theorem:Admissibility of the up fragment})}
\label{section:Proof of theorem:Admissibility of the up fragment}
%%%%%%%%% first case
As a \emph{first case} we show that $\vlrulename{\bvtatiurule}$ is admissible
for $\BVT$. So, we start by assuming:
\[
\vlderivation                           {
\vlin{\bvtatiurule}{}
     {\strS\vlscn{\vlone}}             {
\vlpr{\bvtDder}{}
     {\strS\vlsbr(\atma;\natma)}}}
\]
Applying splitting (Theorem~\ref{theorem:Splitting}) to $\bvtDder$ we have
$\vlstore{\vlsbr[\vlhole;[\strK_1;\strK_2]]}
 \bvtInfer{\bvtDder_0}{\vlfo{\atmb}{{\vlread}} \bvtJ \strS\vlhole}$,
$\vlstore{\vlsbr[\atma;\strK_1]}
 \bvtInfer{\bvtDder_1}{\ \bvtJ {\vlread}} $, and
$\vlstore{\vlsbr[\natma;\strK_2]}
 \bvtInfer{\bvtDder_2}{\ \bvtJ {\vlread}}$, for some
$\strK_1, \strK_2, \atmb$, where $\atmb$, and $\atma$ may coincide.
A basic observation is that $\bvtDder_0$ holds for any structure we may plug
inside $\vlhole$. So, in particular, we have
$\vlstore{\vlsbr[\vlscn{\vlone};\strK_1;\strK_2]}
 \bvtInfer{\bvtDder'_0}
 {\ \vlfo{\atmb}{{\vlread}}
           \bvtJ \strS\vlscn{\vlone}}$.
Now, shallow splitting (Proposition~\ref{proposition:Shallow Splitting}) on
$\bvtDder_1, \bvtDder_2$ implies $\bvtInfer{\bvtPder_1}{\natma \bvtJ {\strK_1}}
$, and $\bvtInfer{\bvtPder_2}{\atma  \bvtJ {\strK_2}}$.
So, we can build the following proof with the same conclusion as $\bvtDder$, but
without its bottommost instance of $\vlrulename{\bvtatiurule}$:
\[
\vlderivation                                        {
\vlde{\bvtDder'_0}{}
     {\strS\vlscn{\vlone}}                           {
\vliq{}{}
     {\vlfo{\atmb}{\vlsbr[\vlscn{\vlone};\strK_1;\strK_2]}}      {
\vlde{\bvtPder_1,\bvtPder_2}{}
     {\vlfo{\atmb}{\vlsbr[\strK_1;\strK_2]}}     {
\vlin{\bvtatidrule}{}
     {\vlfo{\atmb}{\vlsbr[\natma;\atma]}}       {
\vliq{}{}{\vlfo{\atmb}{\vlone}}                {
\vlhy{\vlone}                                  }}}}}}
\]
%%%%%%%%% second case
As a \emph{second case} we show that $\vlrulename{\bvtsequrule}$ is admissible
for $\BVT$. So, we start by assuming:
\[
\vlderivation                                    {
\vlin{\bvtsequrule}{}
     {\strS\vlsbr<(\strR;\strT);(\strU;\strV)>} {
\vlpr{\bvtDder}{}
     {\strS\vlsbr(<\strR;\strT>;<\strU;\strV>) }}}
\]
Applying splitting (Theorem~\ref{theorem:Splitting}) to $\bvtDder$ we have
$\vlstore{\vlsbr[\vlhole;\strK_1;\strK_2]}
 \bvtInfer{\bvtDder_0}
 {\vlfo{\atmb}{{\vlread}} \bvtJ \strS\vlhole}$,
$\vlstore{\vlsbr[<\strR;\strT>;\strK_1]}
 \bvtInfer{\bvtDder_1}{\ \bvtJ {\vlread}}$, and
$\vlstore{\vlsbr[<\strU;\strV>;\strK_2]}
 \bvtInfer{\bvtDder_2}{\ \bvtJ {\vlread}}$, for some
$\strK_1, \strK_2, \atmb$, where $\atmb$, and $\atma$ may coincide.
A basic observation is that $\bvtDder_0$ holds for any structure we may plug
inside $\vlhole$. So, in particular, we have
$\vlstore{\vlsbr[<(\strR;\strT);(\strU;\strV)>;\strK_1;\strK_2]}
 \bvtDder'_0\ :\
          \vlfo{\atmb}{\vlread}
           \bvtJ
           \strS<(\strR;\strT);(\strU;\strV)>$.
Then, shallow splitting (Proposition~\ref{proposition:Shallow Splitting}) on
$\bvtDder_1, \bvtDder_2$ implies
$\vlstore{\vlsbr<\strK_\strR;\strK_\strT>}
 \bvtInfer{\bvtPder_0 }{\vlread \bvtJ {\strK_1}} $,
$\vlstore{\vlsbr{[\strR;\strK_\strR]}}
 \bvtInfer{\bvtPder_1 }{\ \bvtJ {\vlread}}$,
$\vlstore{\vlsbr{[\strT;\strK_\strT]}}
 \bvtInfer{\bvtPder_2 }{\ \bvtJ {\vlread}}$,
$\vlstore{\vlsbr<\strK_\strU;\strK_\strV> \bvtJ {\strK_2}}
 \bvtInfer{\bvtPder'_0}{\vlread}$,
$\vlstore{\vlsbr{[\strU;\strK_\strU]}}
 \bvtInfer{\bvtPder'_1}{\ \bvtJ {\vlread}}$, and
$\vlstore{\vlsbr{[\strV;\strK_\strV]}}
 \bvtInfer{\bvtPder'_2}{\ \bvtJ {\vlread}}$.
So, we can build the following proof with the same conclusion as $\bvtDder$, but
without its bottommost instance of $\vlrulename{\bvtsequrule}$:
\[
\vlderivation                                            {
\vlde{}{}
     {\strS\vlsbr<(\strR;\strT);(\strU;\strV)>}          {
\vlde{\bvtDder'_0}{}
     {\vlfo{\atmb}{\vlsbr[<(\strR;\strT);(\strU;\strV)>
                         ;\strK_1
                         ;\strK_2]  }}                   {
\vlin{\bvtseqdrule}{}
     {\vlfo{\atmb}{\vlsbr[<(\strR;\strT);(\strU;\strV)>
                         ;<\strK_\strR;\strK_\strT>
                         ;<\strK_\strU;\strK_\strV>]}  } {
\vlin{\bvtswirule,\bvtswirule}{}
     {\vlfo{\atmb}{\vlsbr[<[(\strR;\strT);\strK_\strR]
                          ;[(\strU;\strV);\strK_\strT]>
                         ;<\strK_\strU;\strK_\strV>]}  } {
\vlde{}{}
     {\vlfo{\atmb}{\vlsbr[<([\strR;\strK_\strR];\strU)
                          ;([\strT;\strK_\strT];\strV)>
                         ;<\strK_\strU;\strK_\strV>]}}   {
\vliq{}{}
     {\vlfo{\atmb}{\vlsbr[<(\vlone;\strU)
                          ;(\vlone;\strV)>
                         ;<\strK_\strU;\strK_\strV>]}}   {
\vlin{\bvtseqdrule}{}
     {\vlfo{\atmb}{\vlsbr[<\strU;\strV>
                         ;<\strK_\strU;\strK_\strV>]}}   {
\vlde{}{}
     {\vlfo{\atmb}{\vlsbr<[\strU;\strK_\strU]
                         ;[\strV;\strK_\strV]>}}         {
\vliq{}{}
     {\vlfo{\atmb}{\vlsbr<\vlone;\vlone>}}               {
\vlhy{\vlone}                                   }}}}}}}}}}
\]
%%%%%%%%% third case
\par
As a \emph{third case} we show that $\vlrulename{\bvtrurule}$ is admissible for
$\BVT$. So, we start by assuming:
\[
\vlderivation                                               {
\vlin{\bvtrurule}{}
     {\strS\vlfo{\atma}{\vlsbr(\strR;\strT)}}              {
\vlpr{\bvtDder}{}
     {\strS\vlsbr(\vlfo{\atma}{\strR};\vlfo{\atma}{\strT})}}}
\]
Applying splitting (Theorem~\ref{theorem:Splitting}) to $\bvtDder$ we have
$\vlstore{\vlsbr[\vlhole;\strK_1;\strK_2]}
 \bvtInfer{\bvtDder_0}{\vlfo{\atmb}{{\vlread}} \bvtJ \strS\vlhole}$,
$\vlstore{\vlsbr[\vlfo{\atma}{\strR};\strK_1]}
 \bvtInfer{\bvtDder_1}{\ \bvtJ {\vlread}}$,
and
$\vlstore{\vlsbr[\vlfo{\atma}{\strT};\strK_2]}
 \bvtInfer{\bvtDder_2}{\ \bvtJ {\vlread}}$,
for some
$\strK_1, \strK_2, \atmb$, where $\atmb$, and $\atma$ may coincide.
A basic observation is that $\bvtDder_0$ holds for any structure we may plug
inside $\vlhole$. So, in particular, we have
$\vlstore{\vlfo{\atmb}{{\vlsbr[\vlfo{\atma}{\vlsbr(\strR;\strT)}
                       ;\strK_1;\strK_2]}}
          \bvtJ \strS\vlfo{\atma}{\vlsbr(\strR;\strT)}}
 \bvtInfer{\bvtDder'_0}{\ \vlread}$.
Then, shallow splitting (Proposition~\ref{proposition:Shallow Splitting}) on
$\bvtDder_1, \bvtDder_2$ implies
$\bvtInfer{\bvtPder_0 }{\vlfo{\atma}{\strK_\strR} \bvtJ {\strK_1}} $,
$\vlstore{\vlsbr{[\strR;\strK_\strR]}}
 \bvtInfer{\bvtPder_1 }{\ \bvtJ {\vlread}}$, and
$\bvtInfer{\bvtPder'_0 }{\vlfo{\atma}{\strK_\strT} \bvtJ {\strK_2}}$,
$\vlstore{\vlsbr{[\strT;\strK_\strT]}}
 \bvtInfer{\bvtPder'_1 }{\ \bvtJ {\vlread}}$.
So, we can build the following proof with the same conclusion as $\bvtDder$, but
without its bottommost instance of $\vlrulename{\bvtrurule}$:
\[
\vlderivation                                            {
\vlde{\bvtDder'_0}{}
     {\strS\vlfo{\atma}{\vlsbr(\strR;\strT)}}                  {
\vlde{}{}
     {\vlfo{\atmb}{\vlsbr[\vlfo{\atma}
                                {(\strR;\strT)}
                         ;\strK_1
                         ;\strK_2]
                  }}                                     {
\vlin{\bvtrdrule}{}
     {\vlfo{\atmb}{\vlsbr[\vlfo{\atma}
                                {(\strR;\strT)}
                         ;\vlfo{\atma}{\strK_\strR}
                         ;\vlfo{\atma}{\strK_\strT}]
                  }}                                     {
\vlin{\bvtswirule}{}
     {\vlfo{\atmb}{\vlsbr[\vlfo{\atma}
                               {[(\strR;\strT);\strK_\strR]}
                         ;\vlfo{\atma}{\strK_\strT}]
                  }}                                     {
\vlin{\bvtrdrule}{}
     {\vlfo{\atmb}{\vlsbr[\vlfo{\atma}
                               {([\strR;\strK_\strR];\strT)}
                         ;\vlfo{\atma}{\strK_\strT}]
                  }}                                     {
\vliq{\bvtswirule}{}
     {\vlfo{\atmb}{\vlfo{\atma}
                  {\vlsbr[([\strR;\strK_\strR];\strT)
                         ;\strK_\strT]
                         }}}                             {
\vlde{}{}
     {\vlfo{\atmb}{\vlfo{\atma}
                  {\vlsbr([\strT;\strK_\strT]
                         ;[\strR;\strK_\strR])}}}        {
\vliq{}{}
     {\vlfo{\atmb}{\vlfo{\atma}{\vlsbr(\vlone;\vlone)}}} {
\vlhy{\vlone}                                    }}}}}}}}}
\]
\section{Proof of \textit{Internalizing sequents} 
(Proposition~\ref{proposition:Internalizing sequents},
page~\pageref{proposition:Internalizing sequents})}
\label{section:Proof of Internalizing sequents}
By induction on the size $\Size{\mllPder}$ of $\mllPder$ which counts the
number of instance rules in it, proceeding by cases on its last rule. To avoid
cluttering the derivations of $\SBV$ we are going to produce, we shall omit
$\mapMLLToBVT{(\cdot)}$ around the formulas.
%%%%% \mllax
\par
Let the last rule of $\mllPder$ be $\vlrulename{\mllax}$. Then, the derivation
we are looking for is the structure $\vlsbr<\mllnA;\mllA>$.
%%%%% \mllcut
\par
Let the last rule of $\mllPder$ be $\vlrulename{\mllcut}$. Then:
\[
\vlderivation                             {
\vliq{\bvtinturule}{}
     {\vlsbr
      <(\vlne{\mllA_1};\vldots;\vlne{\mllA_m}
        ;\vlne{\mllB_1};\vldots;\vlne{\mllB_m}
       )
       ;\mllB_m>
     }                                    {
\vliq{}{}
     {\vlsbr
      <(<(\vlne{\mllA_1};\vldots;\vlne{\mllA_m});(\mllnA;\mllA)>
        ;\vlne{\mllB_1};\vldots;\vlne{\mllB_m}
       )
       ;\mllB_m>
     }                                    {
\vlin{\bvtsequrule}{}
     {\vlsbr
      <(<(\vlone;(\vlne{\mllA_1};\vldots;\vlne{\mllA_m}));(\mllnA;\mllA)>
        ;\vlne{\mllB_1};\vldots;\vlne{\mllB_m}
       )
       ;\mllB_m>
     }                                    {
\vliq{}{}
     {\vlsbr
      <((<\vlone;\mllnA>
         ;<(\vlne{\mllA_1};\vldots;\vlne{\mllA_m});\mllA>
        );\vlne{\mllB_1};\vldots;\vlne{\mllB_m})
       ;\mllB_m>
     }                                    {
\vlin{\bvtsequrule}{}
     {\vlsbr
      <(<(\vlne{\mllA_1};\vldots;\vlne{\mllA_m});\mllA>
        ;(\mllnA;\vlne{\mllB_1};\vldots;\vlne{\mllB_m}))
       ;
       (\vlone;\mllB_m)>
     }                                    {
\vliq{}{}
     {\vlsbr
      (<<(\vlne{\mllA_1};\vldots;\vlne{\mllA_m});\mllA>;\vlone>
       ;
       <(\mllnA;\vlne{\mllB_1};\vldots;\vlne{\mllB_m});\mllB_m>)
     }                                    {
\vlde{\vlsbr(\mapMLLToBVT{\mllPderA};\mapMLLToBVT{\mllPderB})}{}
     {\vlsbr
      (<(\vlne{\mllA_1};\vldots;\vlne{\mllA_m});\mllA>
       ;
       <(\mllnA;\vlne{\mllB_1};\vldots;\vlne{\mllB_m});\mllB_m>)
     }                                    {
\vlhy{\vlsbr
      ((<\vlne{\mllA_1};\mllA_1>;\vldots;<\vlne{\mllA_m};\mllA_m>
                                        ;<\mllnA;\mllA>)
       ;
       (<\mllnA;\mllA>
        ;<\vlne{\mllB_1};\mllB_1>;\vldots;<\vlne{\mllB_m};\mllB_m>))
     }                                    }}}}}}}}
\]
exists under the inductive hypothesis that $\mllPderA, \mllPderB$ derive the
assumptions of $\mllcut$.
%%%%% \mlltens
\par
Let the last rule of $\mllPder$ be $\vlrulename{\mlltens}$. Then, there is:
\[
\vlderivation                             {
\vlin{\bvtseqdrule}{}
     {\vlsbr
      <(\vlne{\mllA_1};\vldots;\vlne{\mllA_m}
       ;\vlne{\mllB_1};\vldots;\vlne{\mllB_n})
      ;(\mllA;\mllB)
      >
     }                                    {
\vlde{\vlsbr(\mapMLLToBVT{\mllPderA}
             ;
             \mapMLLToBVT{\mllPderB}
            )}{}
     {\vlsbr
      (<(\vlne{\mllA_1};\vldots;\vlne{\mllA_m});\mllA>
      ;<(\vlne{\mllB_1};\vldots;\vlne{\mllB_n});\mllB>
      )
     }                                    {
\vlhy{\vlsbr
      ((<\vlne{\mllA_1};\mllA_1>
         ;\vldots;
        <\vlne{\mllA_m};\mllA_m>
       )
       ;
       (<\vlne{\mllB_1};\mllB_1>
         ;\vldots;
        <\vlne{\mllB_n};\mllB_n>
       )
      )
     }                                    }}}
\]
under the inductive hypothesis that $\mllPderA, \mllPderB$ derive the
assumptions of
$\mlltens$.
%%%%% \mllimpl
\par
Let the last rule of $\mllPder$ be $\vlrulename{\mllimpl}$. Then:
\[
\vlderivation                             {
\vlin{\bvtpmixrule}{}
     {\vlsbr
      <(\vlne{\mllA_1};\vldots;\vlne{\mllA_m});
       [\vlne{\mllA};\mllB]>
     }                                    {
\vlin{\bvtmixprule}{}
     {\vlsbr
      <(\vlne{\mllA_1};\vldots;\vlne{\mllA_m});
       <\vlne{\mllA};\mllB>>
     }                                    {
\vliq{}{}
     {\vlsbr
      ((\vlne{\mllA_1};\vldots;\vlne{\mllA_m});
        <\vlne{\mllA};\mllB>)
     }                                    {
\vlin{\bvtseqdrule}{}
     {\vlsbr
      (<(\vlne{\mllA_1};\vldots;\vlne{\mllA_m});\vlone>;
       <\vlne{\mllA};\mllB>)
     }                                    {
\vliq{}{}
     {\vlsbr
      <((\vlne{\mllA_1}
         ;\vldots;
         \vlne{\mllA_m});\vlne{\mllA});(\vlone;\mllB)>
     }                                    {
\vlde{\mapMLLToBVT{\mllPder}}{}
     {\vlsbr
      <(\vlne{\mllA_1}
       ;\vldots;
       \vlne{\mllA_m};\vlne{\mllA});\mllB>
     }                                    {
\vlhy{\vlsbr
       (<\vlne{\mllA_1};\mllA_1>
        ;\vldots;
        <\vlne{\mllA_m};\mllA_m>;<\vlne{\mllA};\mllA>)
     }                                    }}}}}}}
\]
exists under the inductive hypothesis that $\mllPder$ derives the assumption of
$\mllimpl$.

\section{Proof of \textit{Output renaming} (Lemma~\ref{lemma:Output renaming},
page~\pageref{lemma:Output renaming})}
\label{section:Proof of lemma:Output renaming}
%%%%%%%% base case
Let $\llcxM\equiv\llcxX$. Then:
\[
\vlderivation                                            {
\vlin{\bvtdefdrule}{}
     {\vlsbr[<\llcxX;\llcxnp>
            ;(\llcxp;\llcxno)]\equiv
      \vlsbr[\mapLcToDi{\llcxX}{\llcxp}
            ;(\llcxp;\llcxno)]}                          {
\vlhy{\mapLcToDi{\llcxX}{\llcxo}\equiv
      \vlsbr<\llcxX;\llcxno>}                            }}
\]
%%%%%%%% first case
\par
Let $\llcxM\equiv\llcxA{\llcxP}{\llcxQ}$. Then:
\[
\vlderivation                                            {
\vliq{}{}
     {\vlsbr[\vlex{\llcxpP}
                  {[\mapLcToDi{\llcxP}{\llcxpP}
                   ;\vlex{\llcxq}{\mapLcToDi{\llcxQ}{\llcxq}}
                   ;(\llcxpP;\llcxnp)
                   ]};(\llcxp;\llcxno)]
      \equiv\vlsbr[\mapLcToDi{\llcxA{\llcxP}{\llcxQ}}
                             {\llcxo};(\llcxp;\llcxno)]   }   {
\vlin{\bvtrdrule}{}
     {\vlsbr[\vlex{\llcxpP}
                  {[\mapLcToDi{\llcxP}{\llcxpP}
                   ;\vlex{\llcxq}{\mapLcToDi{\llcxQ}{\llcxq}}
                   ;(\llcxpP;\llcxnp)
                   ]};\vlex{\llcxpP}{(\llcxp;\llcxno)}]   }   {
\vliq{\bvtswirule}{}
     {\vlex{\llcxpP}
           {\vlsbr[\mapLcToDi{\llcxP}{\llcxpP}
                  ;\vlex{\llcxq}{\mapLcToDi{\llcxQ}{\llcxq}}
                  ;(\llcxpP;\llcxnp);(\llcxp;\llcxno)]}   }   {
\vliq{\bvtswirule}{}
     {\vlex{\llcxpP}
           {\vlsbr[\mapLcToDi{\llcxP}{\llcxpP}
                  ;\vlex{\llcxq}{\mapLcToDi{\llcxQ}{\llcxq}}
                  ;([\llcxnp;(\llcxp;\llcxno)];\llcxpP)]}   }  {
\vliq{\bvtatidrule}{}
     {\vlex{\llcxpP}
           {\vlsbr[\mapLcToDi{\llcxP}{\llcxpP}
                  ;\vlex{\llcxq}{\mapLcToDi{\llcxQ}{\llcxq}}
                  ;(\llcxno;[\llcxnp;\llcxp];\llcxpP)]}   }  {
\vlhy{\mapLcToDi{\llcxA{\llcxP}{\llcxQ}}{\llcxo}\equiv
      \vlex{\llcxpP}
           {\vlsbr[\mapLcToDi{\llcxP}{\llcxpP}
                  ;\vlex{\llcxq}{\mapLcToDi{\llcxQ}{\llcxq}}
                  ;(\llcxpP;\llcxno)]}}                  }}}}}}
\]
%%%%%%%% second case
\par
Let $\llcxM\equiv\llcxF{\llcxX}{\llcxP}$. Then:
\[
\vlderivation                                            {
\vliq{}{}
     {\vlsbr[\vlfo{\llcxX}{\vlex{\llcxpP}
                          {[\mapLcToDi{\llcxP}{\llcxpP}
                           ;(\llcxpP;\llcxnp)
                           ]}};(\llcxp;\llcxno)]
      \equiv\vlsbr[\mapLcToDi{\llcxF{\llcxX}{\llcxP}}
                             {\llcxo};(\llcxp;\llcxno)]   }   {
\vliq{\bvtrdrule,\bvtrdrule}{}
     {\vlsbr[\vlfo{\llcxX}{\vlex{\llcxpP}
                          {[\mapLcToDi{\llcxP}{\llcxpP}
                           ;(\llcxpP;\llcxnp)
                           ]}};\vlex{\llcxX}
                                    {\vlex{\llcxpP}{(\llcxp;\llcxno)}}]}   {
\vliq{\bvtswirule,\bvtatidrule,\bvtpmixrule,\bvtmixprule}{}
     {\vlfo{\llcxX}{\vlex{\llcxpP}
                          {\vlsbr[\mapLcToDi{\llcxP}{\llcxpP}
                                 ;(\llcxpP;\llcxnp)
                                 ;(\llcxp;\llcxno)]}}   }  {
\vlhy{\mapLcToDi{\llcxF{\llcxX}{\llcxP}}{\llcxo}\equiv
      \vlfo{\llcxX}{\vlex{\llcxpP}
                          {\vlsbr[\mapLcToDi{\llcxP}{\llcxpP}
                                 ;(\llcxpP;\llcxno)]}} }}}}}
\]
%%%%%%%% third case
\par
Let $\llcxM\equiv\llcxE{\llcxP}{\llcxX}{\llcxQ}$. Then:
\[
\vlderivation                                              {
\vliq{}{}
     {\vlsbr[\vlfo{\llcxX}{[\mapLcToDi{\llcxP}{\llcxp}
                           ;\mapLcToDi{\llcxQ}{\llcxX}
                           ]};(\llcxp;\llcxno)]
      \equiv\vlsbr[\mapLcToDi{\llcxE{\llcxP}{\llcxX}{\llcxQ}}
                             {\llcxo};(\llcxp;\llcxno)]}   {
\vliq{\bvtrdrule}{}
     {\vlsbr[\vlfo{\llcxX}{[\mapLcToDi{\llcxP}{\llcxp}
                           ;\mapLcToDi{\llcxQ}{\llcxX}
                           ]};\vlex{\llcxX}
                                   {(\llcxp;\llcxno)}] }   {
\vlde{\bvtDder}{}
     {\vlfo{\llcxX}{\vlsbr[\mapLcToDi{\llcxP}{\llcxp}
                           ;(\llcxp;\llcxno)
                           ;\mapLcToDi{\llcxQ}{\llcxX}]}}  {
\vlhy{\mapLcToDi{\llcxE{\llcxP}{\llcxX}{\llcxQ}}
                              {\llcxo}
      \equiv
      \vlfo{\llcxX}{\vlsbr[\mapLcToDi{\llcxP}{\llcxo}
                          ;\mapLcToDi{\llcxQ}{\llcxX}]} }}}}}
\]
where $\bvtDder$ exists thanks to the inductive hypothesis which holds
because $\llcxP$ is a sub-term of $\llcxA{\llcxP}{\llcxQ}$.

\section{Proof of \textit{Simulating $\llcxBeta$} (Lemma~\ref{lemma:Simulating
llcxBeta}, page~\pageref{lemma:Simulating llcxBeta})}
\label{section:lemma:Simulating llcxBeta}
%%%%% \bvtsvarule
Let us focus on $\vlrulename{\bvtsvarule}$. The following derivation exists:
\[
\vlderivation                                              {
\vliq{}{}
     {\mapLcToDi{\llcxE{\llcxX}{\llcxX}{\llcxP}}{\atmo}
      \equiv
      \vlfo{\llcxX}
           {\vlsbr[\mapLcToDi{\llcxX}{\llcxo}
                  ;\mapLcToDi{\llcxP}{\llcxX}
                  ]}
      \equiv
      \vlfo{\llcxX}
           {\vlsbr[<\llcxX;\llcxno>
                  ;\mapLcToDi{\llcxP}{\llcxX}
                  ]}
     }                                                     {
\vlin{\bvtmixprule}{}
     {
      \vlfo{\llcxX}
           {\vlsbr[\mapLcToDi{\llcxP}{\llcxX}
                  ;<\llcxX;\llcxno>
                  ]}
     }                                                     {
\vlde{\bvtDder}{}
     {
      \vlfo{\llcxX}
           {\vlsbr[\mapLcToDi{\llcxP}{\llcxX}
                  ;(\llcxX;\llcxno)
                  ]}
     }                                                     {
\vliq{}{}
     {\vlfo{\llcxX}{\mapLcToDi{\llcxP}{\atmo}}}            {
\vlhy{\mapLcToDi{\llcxP}{\atmo}}                           }}}}}
\]
\noindent
where $\bvtDder$ exists thanks to Lemma~\ref{lemma:Output renaming}. The here
above derivation requires $\vlrulename{\bvtsequrule}$ because
$\vlrulename{\bvtmixprule}$ is derivable in $\Set{\vlrulename{\bvtsequrule}}$.
%%%%% \bvtsinrule
\par
Let us focus on $\vlrulename{\bvtsinrule}$. The following derivation exists:
\[
\vlderivation                                              {
\vliq{\mathtt{e}_0}{}
     {\mapLcToDi{\llcxA{\llcxF{\llcxX}{\llcxM}}{\llcxN}}{\llcxo}
      \equiv
      \vlex{\llcxp}
           {\vlsbr[\vlfo{\llcxX}
			{\vlex{\llcxp'}
			      {[\mapLcToDi{\llcxM}{\llcxp'}
			       ;(\llcxp';\llcxnp)]}
			}
                  ;\vlex{\llcxq}{\mapLcToDi{\llcxN}{\llcxq}}
                  ;(\llcxp;\llcxno)]
           }
     }                                                     {
\vlin{\bvtrdrule}{}
     {\vlex{\llcxp}
           {\vlsbr[\vlfo{\llcxX}
			{\vlex{\llcxp'}
			      {[\mapLcToDi{\llcxM}{\llcxp'}
			       ;(\llcxp';\llcxnp)]
                              }
			}
                  ;\vlex{\llcxX}{\mapLcToDi{\llcxN}{\llcxX}}
                  ;(\llcxp;\llcxno)]
            }
     }                                                     {
\vlde{\bvtDder}{}
     {
      \vlex{\llcxp}
           {\vlsbr[\vlfo{\llcxX}
			{[\vlex{\llcxp'}
			      {[\mapLcToDi{\llcxM}{\llcxp'}
			       ;(\llcxp';\llcxnp)]
                              }
                         ;\mapLcToDi{\llcxN}{\llcxX}]
                        }
                  ;(\llcxp;\llcxno)]
            }
     }                                                     {
\vliq{\mathtt{e}_1}{}
     {
      \vlex{\llcxp}
           {\vlsbr[\vlfo{\llcxX}
			{[\vlex{\llcxp'}
			      {\mapLcToDi{\llcxM}{\llcxp}
                              }
                         ;\mapLcToDi{\llcxN}{\llcxX}]
                        }
                  ;(\llcxp;\llcxno)]
            }
     }                                                     {
\vlde{\bvtDder'}{}
     {\vlex{\llcxp}
           {\vlsbr[\vlfo{\llcxX}
			{[\mapLcToDi{\llcxM}{\llcxp}
                         ;\mapLcToDi{\llcxN}{\llcxX}]
                        }
                  ;(\llcxp;\llcxno)]
            }
      \equiv
      \vlex{\llcxp}
           {\vlsbr[\mapLcToDi{\llcxE{\llcxM}{\llcxX}{\llcxN}}{\llcxp}
                  ;(\llcxp;\llcxno)]
           }
     }                                                     {
\vliq{\mathtt{e}_2}{}
     {\vlex{\llcxp}
           {\mapLcToDi{\llcxE{\llcxM}{\llcxX}{\llcxN}}{\llcxo}
           }
     }                                                     {
\vlhy{\mapLcToDi{\llcxE{\llcxM}{\llcxX}{\llcxN}}{\llcxo}}}}}}}}}
\]
where:
\begin{itemize}
\item
$\vlex{\llcxq}{\mapLcToDi{\llcxN}{\llcxq}}$ in the conclusion of $\mathtt{e}_0$
becomes
$\vlex{\llcxX}
      {\mapLcToDi{\llcxN}{\llcxq}\subst{\llcxX}{\llcxq}}
 \approx
 \vlex{\llcxX}{\mapLcToDi{\llcxN}
      {\llcxX}}$
in its premise because $\llcxq$ only occurs as output channel name in a pair
$\vlsbr(\llcxp'';\llcxq)$, for some $\llcxp''$, and nowherelse;
\item
Lemma~\ref{lemma:Output renaming} implies the existence of both
$\bvtDder, \bvtDder'$;
\item
in the conclusion of $\mathtt{e}_1$, $\llcxp'$ has disappeared from
$\mapLcToDi{\llcxM}{\llcxp}$;
\item
in the conclusion of $\mathtt{e}_2$, $\llcxp$ has disappeared from
$\mapLcToDi{\llcxE{\llcxM}{\llcxX}{\llcxN}}{\llcxo}$.
\end{itemize}
%%%%% \bvtslrule
\par
Let us focus on $\vlrulename{\bvtslrule}$. The following derivation exists:
\[
\vlderivation                                              {
\vliq{\mathtt{e}_0}{}
     {\mapLcToDi{\llcxE{\llcxF{\llcxY}{\llcxM}}{\llcxX}{\llcxP}}
                {\atmo}
      \equiv
      \vlfo{\llcxX}
           {\vlsbr[\vlfo{\llcxY}
                        {\vlex{\llcxp}
                              {[\mapLcToDi{\llcxM}{\llcxp}
                               ;(\llcxp;\llcxno)]}
                        }
                   ;\mapLcToDi{\llcxP}{\llcxX}
                  ]
           }
     }                                                     {
\vlin{\bvtrdrule,\bvtrdrule}{}
     {
      \vlfo{\llcxX}
           {\vlsbr[\vlfo{\llcxY}
                        {\vlex{\llcxp}
                              {[\mapLcToDi{\llcxM}{\llcxp}
                               ;(\llcxp;\llcxno)]}
                        }
                   ;\vlex{\llcxY}
                         {\vlex{\llcxp}
                               {\mapLcToDi{\llcxP}{\llcxX}}}
                  ]
           }
     }                                                     {
\vliq{\mathtt{e}_1}{}
     {
      \vlfo{\llcxX}
           {\vlfo{\llcxY}
                 {\vlex{\llcxp}
                       {\vlsbr[\mapLcToDi{\llcxM}{\llcxp}
                               ;(\llcxp;\llcxno)
                               ;\mapLcToDi{\llcxP}{\llcxX}
                              ]
                       }
                 }
           }
     }                                                     {
\vlhy{\vlfo{\llcxY}
           {\vlex{\llcxp}
                 {\vlsbr[\vlfo{\llcxX}
                              {[\mapLcToDi{\llcxM}{\llcxp}
                               ;\mapLcToDi{\llcxP}{\llcxX}
                               ]
                              }
                         ;(\llcxp;\llcxno)
                        ]
                 }
           }
     \equiv
      \vlfo{\llcxY}
           {\vlex{\llcxp}
                 {\vlsbr[\mapLcToDi{\llcxE{\llcxM}{\llcxX}{\llcxP}}{\llcxp}
                         ;(\llcxp;\llcxno)
                        ]
                 }
           }
     \equiv
     \mapLcToDi{\llcxF{\llcxY}{\llcxE{\llcxM}{\llcxX}{\llcxP}}}
               {\atmo}
       }}}}}
\]
where $\llcxp$, and $\llcxY$ do not belong to $\mapLcToDi{\llcxP}{\llcxX}$, and
$\mathtt{e}_1$ applies three of the axioms in Figure~\ref{fig:BVT-structure
equivalence}.
%%%%% \bvtsalrule
\par
Let us focus on $\vlrulename{\bvtsalrule}$. The following derivation
exists:
\[
\vlderivation                                              {
\vliq{}{}
     {\mapLcToDi{\llcxE{\llcxA{\llcxM}{\llcxN}}{\llcxX}{\llcxP}}
                {\atmo}
      \equiv
      \vlfo{\llcxX}
           {\vlsbr[\mapLcToDi{\llcxA{\llcxM}{\llcxN}}
                             {\llcxo}
                  ;\mapLcToDi{\llcxP}{\llcxX}]
           }
      \equiv
      \vlfo{\llcxX}
           {\vlsbr[\vlex{\llcxp}
                        {[\mapLcToDi{\llcxM}
                                    {\llcxp}
                          ;\vlex{\llcxq}{\mapLcToDi{\llcxN}{\llcxq}}
                          ;(\llcxp;\llcxno)
                         ]
                        }
                   ;\mapLcToDi{\llcxP}{\llcxX}
                  ]
           }
     }                                                     {
\vlin{\bvtrdrule}{}
     {
      \vlfo{\llcxX}
           {\vlsbr[\vlex{\llcxp}
                        {[\mapLcToDi{\llcxM}
                                    {\llcxp}
                          ;\vlex{\llcxq}{\mapLcToDi{\llcxN}{\llcxq}}
                          ;(\llcxp;\llcxno)
                         ]
                        }
                   ;\vlex{\llcxp}{\mapLcToDi{\llcxP}{\llcxX}}
                  ]
           }
     }                                                     {
\vliq{}{}
     {
      \vlfo{\llcxX}
           {\vlex{\llcxp}
                        {\vlsbr
                         [\mapLcToDi{\llcxM}
                                    {\llcxp}
                          ;\vlex{\llcxq}{\mapLcToDi{\llcxN}{\llcxq}}
                          ;(\llcxp;\llcxno)
                          ;\mapLcToDi{\llcxP}{\llcxX}
                         ]
                        }
           }
     }                                                     {
\vliq{}{}
     {
      \vlex{\llcxp}
           {\vlfo{\llcxX}
                        {\vlsbr
                         [\mapLcToDi{\llcxM}
                                    {\llcxp}
                          ;\vlex{\llcxq}{\mapLcToDi{\llcxN}{\llcxq}}
                          ;(\llcxp;\llcxno)
                          ;\mapLcToDi{\llcxP}{\llcxX}
                         ]
                        }
           }
     }                                                     {
\vlhy{\begin{array}{l}
      \mapLcToDi{\llcxA{\llcxE{\llcxM}{\llcxX}{\llcxP}}
                                      {\llcxN}
                }
                {\llcxo}
       \\
       \equiv
       \vlex{\llcxp}
            {\vlsbr[\mapLcToDi{\llcxE{\llcxM}{\llcxX}{\llcxP}}
                              {\llcxp}
                    ;\vlex{\llcxq}{\mapLcToDi{\llcxN}{\llcxq}}
                    ;(\llcxp;\llcxno)
                   ]
            }
       \\
       \equiv
      \vlex{\llcxp}
           {\vlsbr[\vlfo{\llcxX}
                        {[\mapLcToDi{\llcxM}
                                   {\llcxp}
                          ;\mapLcToDi{\llcxP}{\llcxX}
                         ]
                        }
                   ;\vlex{\llcxq}{\mapLcToDi{\llcxN}{\llcxq}}
                   ;(\llcxp;\llcxno)
                  ]
           }
      \end{array}
       }}}}}}
\]
%%%%% \bvtsarrule
The case relative to $\vlrulename{\bvtsarrule}$ develops as for
$\vlrulename{\bvtsalrule}$.

\section{Proof of \textit{\textbf{Completeness of $\SBVT$}}
(Proposition~\ref{theorem:Completeness of SBVT},
page~\pageref{theorem:Completeness of SBVT})}
\label{section:Proof of Completeness of SBVT}
By induction on $\Size{\llcxSOSJud{\llcxM}{\llcxN}}$, proceeding by cases on the
last rule used, taken among those in Figure~\ref{fig:Rewriting relation
llcxSOSJud}.
%%%%% \llcxSOSBrule
\par
Let the last rule be $\vlrulename{\llcxSOSBrule}$, namely
$\llcxSOSJud{\llcxM}{\llcxN}$ because $\llcxM \llcxBeta \llcxN$.
Lemma~\ref{lemma:Simulating llcxBeta} directly implies the thesis.
%%%%% \llcxSOSslrule
\par
Let the last rule be $\vlrulename{\llcxSOStrarule}$. The inductive hypothesis
implies the existence of $\bvtDder_0, \bvtDder_1$:
\[
\vlderivation                                                   {
\vlde{\bvtDder_0}{}
     {\mapLcToDi{\llcxM}{\atmo}
     }                                                           {
\vlde{\bvtDder_1}{}
     {\mapLcToDi{\llcxP}{\atmo}
     }                                                           {
\vlhy{\mapLcToDi{\llcxN}{\atmo}
     }                                                            }}}
\]
%%%%% \llcxSOSslrule
\par
Let the last rule be $\vlrulename{\llcxSOSsrrule}$. The inductive hypothesis
implies the existence of $\bvtDder$:
\[
\vlderivation                                    {
\vlde{\bvtDder}{}
     {\mapLcToDi{\llcxE{\llcxP}{\llcxX}{\llcxM}}
                {\atmo}
      \equiv
      \vlfo{\llcxX}
           {\vlsbr[\mapLcToDi{\llcxP}{\llcxo}
                  ;\mapLcToDi{\llcxM}{\llcxX}]}
     }                                           {
\vlhy{\vlfo{\llcxX}
           {\vlsbr[\mapLcToDi{\llcxP}{\llcxo}
                  ;\mapLcToDi{\llcxN}{\llcxX}]}
      \equiv
      \mapLcToDi{\llcxE{\llcxP}{\llcxX}{\llcxN}}
                {\atmo}
     }                                           }}
\]
In all the remaining cases we can proceed just as here above.

\end{document}